\documentclass[11pt]{article}

\usepackage{latexsym,amsmath,amscd,amssymb,graphics,amsfonts}
\usepackage{enumerate}

\usepackage{graphicx}

\usepackage[square,authoryear]{natbib}
\usepackage{url}
\usepackage{colordvi}
\usepackage{eucal}

\usepackage[all]{xy}

\makeatletter

\@addtoreset{figure}{section}
\def\thefigure{\thesection.\@arabic\c@figure}
\def\fps@figure{h, t}
\@addtoreset{table}{bsection}
\def\thetable{\thesection.\@arabic\c@table}
\def\fps@table{h, t}
\@addtoreset{equation}{section}

\makeatother

\begin{document}

\newtheorem{theorem}{Theorem}[section]
\newtheorem{definition}[theorem]{Definition}
\newtheorem{lemma}[theorem]{Lemma}
\newtheorem{remark}[theorem]{Remark}
\newtheorem{proposition}[theorem]{Proposition}
\newtheorem{corollary}[theorem]{Corollary}
\newtheorem{example}[theorem]{Example}
\bibliographystyle{plainnat}

\def\below#1#2{\mathrel{\mathop{#1}\limits_{#2}}}



\title{A new Lagrangian dynamic reduction in field theory}
\author{Fran\c{c}ois Gay-Balmaz$^{1}$ and Tudor S. Ratiu$^{1}$}
\addtocounter{footnote}{1} \footnotetext{Section de
Math\'ematiques and Bernoulli Center, \'Ecole Polytechnique F\'ed\'erale de Lausanne.
CH--1015 Lausanne. Switzerland. Partially supported by a Swiss NSF grant.
\texttt{Francois.Gay-Balmaz@epfl.ch, Tudor.Ratiu@epfl.ch}
\addtocounter{footnote}{1} }


\date{ }
\maketitle

\makeatother
\maketitle

\begin{center}\large
\noindent Annales de l'Institut Fourier, \textbf{16}(3), (2010), 1125--1160.
\end{center}

\noindent \textbf{AMS Classification:} 70S05, 70S15, 70H03, 70H30, 70H99, 58E30, 58E40

\noindent \textbf{Keywords:} covariant reduction, dynamic reduction, affine Euler-Poincar\'e equation, covariant Euler-Poincar\'e equation, Lagrangian, principal bundle field theory 

\begin{abstract}
For symmetric classical field theories on principal bundles there are two methods of symmetry reduction: covariant and dynamic. Assume that the classical field theory is given by a symmetric covariant Lagrangian density defined on the first jet bundle of a principal bundle. It is shown that 
covariant and dynamic reduction  lead to equivalent equations of motion.  This is achieved by constructing a new Lagrangian defined on an infinite dimensional space which turns out to be gauge group invariant.
\end{abstract}


\section{Introduction}\label{introduction}

The solution of an evolutionary partial differential equation can be viewed either as a curve in an infinite dimensional space or as a section of a bundle over spacetime. These two points of view lead to the two classical geometric approaches for the study of these partial differential equations and lead either to dynamical systems in infinite dimensions or to the evolution of sections in jet bundles of finite dimensional manifolds.

In the case of conservative evolutionary equations there is additional structure that plays a fundamental role. Such equations have  Lagrangian and Hamiltonian formulations. Regarding the solutions as curves in infinite dimensional spaces leads to the \textit{dynamic} point of view which presents the equations as Lagrangian or Hamiltonian vector fields on the tangent and cotangent bundles of an infinite dimensional configuration manifold, respectively. In the Lagrangian formulation, the equations are obtained from the usual variational principle for a real valued Lagrangian function on the 
tangent bundle.
Interpreting the solutions as sections of a bundle over spacetime leads to the \textit{covariant} point of view. In this approach, the equations appear as the Euler-Lagrange equations of a Lagrangian density defined on the first jet bundle of a given fiber bundle with values in the densities over spacetime. It is well known that the covariant approach admits a dynamic formulation by choosing a slicing of the configuration bundle over spacetime (see e.g. \cite{GoIsMaMo2004}, \cite{GoIsMa2004} and references therein for the details of this construction).

Assume that a field theoretical Lagrangian density admits a Lie group of symmetries. Then covariant reduction can be performed if the configuration bundle is principal  and the symmetry is a subgroup of the structure group (see \cite{CaRaSh2000}, \cite{CaGaRa2001}, \cite{CaRa2003}). The theory for general covariant Lagrangian reduction is still in development and the principal bundle case mentioned before is the only one which is completely understood. In the dynamic approach, if the Lagrangian has a symmetry group, one can apply the usual reduction procedures from classical mechanics. Thus, given a symmetric field theoretical Lagrangian density, the question naturally arises if one can construct an associated dynamic Lagrangian admitting also a symmetry and if yes, how are the two reduction procedures (covariant and dynamic) related. 

In this paper we propose a new approach to dynamic reduction of classical field theories on principal bundles. Let $P \rightarrow X$ be a right principal $G$-bundle and $\mathcal{L}: J ^1P \rightarrow \Lambda^{n+1}X $ a  Lagrangian density, where $\dim X = n+1$. For simplicity, we assume that the spacetime  $X$ admits a standard slicing $X = \mathbb{R}\times M $  for some manifold $M $ and that $P  = \mathbb{R} \times P_M $, where  $P_M \rightarrow M $ is a principal $G $-bundle. To the density $\mathcal{L}$ we associate a family of time-dependent Lagrangians $L_{\Gamma _0}^\mathcal{L}: T \mathcal{G}au(P_M) \rightarrow \mathbb{R}$,  where $\mathcal{G}au(P_M)$ is the gauge group of the $G $-principal bundle $P _M \rightarrow M $ and $\Gamma_0 $ is a connection on $P_M $. This gives a function $L^\mathcal{L}:T \mathcal{G}au(P_M) \times \mathcal{C}onn(P_M) \rightarrow \mathbb{R}$ if we think of $\Gamma_0 $ as a variable; here $\mathcal{C}onn(P_M)$ denotes the space of connections on the bundle $P_M \rightarrow M $. Remarkably, if $\mathcal{L}$ is $G$-invariant, it turns out that  $L^\mathcal{L}$ is $\mathcal{G}au(P_M)$-invariant. This is reminiscent of the Utiyama trick by which one defines a gauge invariant Lagrangian from a $G$-invariant Lagrangian by substituting the derivatives by covariant derivatives. 
Thus the affine Euler-Poincar\'e reduction, introduced in \cite{GBRa2008b} in order to deal with symmetry reduction for complex fluids, can be applied to $L ^\mathcal{L}$ to yield a reduced Lagrangian $l^\mathcal{L}: \mathfrak{gau}(P_M) \times \mathcal{C}onn(P_M) \rightarrow \mathbb{R}$ together with the associated reduced equations of motion and variational principle on $\mathfrak{gau}(P_M) \times \mathcal{C}onn(P_M) $, where  $\mathfrak{gau}(P_M)$ denotes the Lie algebra of 
$\mathcal{G}au(P_M) $. In this context, the reduced Euler-Lagrange equations are called \textit{affine Euler-Poincar\'e equations}. Concerning the covariant description, since the Lagrangian density $\mathcal{L}$ is $G$-invariant,  one can perform covariant reduction leading to a Lagrangian density $\ell: (J ^1P)/G \rightarrow \Lambda^{n+1}X$ together with the equations of motion and variational principle on $(J ^1P)/G$. In this context, the reduced Euler-Lagrange equations are called \textit{covariant Euler-Poincar\'e equations}. The main result of the paper states that the equations obtained by covariant reduction of the density $\mathcal{L}$ are equivalent to the affine Euler-Poincar\'e equations obtained by dynamical reduction of the Lagrangian $L^\mathcal{L}$. Thus, our results show that the underlying geometry of this new dynamical reduction  is identical to the one usually employed in rigid body dynamics or continuum mechanics. 

It should be noted that there is a main difference between the covariant and the dynamic reduction processes. A solution of the covariantly reduced system, that is, a connection $\sigma$ on $P \rightarrow X $, is the projection of a solution of the original system if and only if $\sigma$ is flat. In the dynamic approach no additional condition is imposed to reconstruct solutions. It is shown that the advection equation in the affine Euler-Poincar\'e system in the dynamic approach is equivalent to the flatness of the connection, which is the compatibility condition for reconstruction in the covariant approach. 

We note that our construction is different from the one in \cite{CaMa2008}. 
They associate to $\mathcal{L}$ a time dependent \textit{instantaneous Lagrangian} $L: T \Gamma(P_M) \rightarrow \mathbb{R}$ obtained by the classical procedure (see e.g. \cite{GoIsMaMo2004}, \cite{GoIsMa2004}), where $\Gamma(P_M)$ denotes the space of local sections of the principal $G $-bundle $P_M \rightarrow M $. If $\mathcal{L}$ is $G $-invariant then the construction of $L $ implies that it is also $G $-invariant and hence, by Lagrange-Poincar\'e reduction, one obtains a reduced Lagrangian $l: (T \Gamma(P_M))/G \rightarrow \mathbb{R}$ and the corresponding Lagrange-Poincar\'e equations (see \cite{CeMaRa2001} for details of this construction). In \cite{CaMa2008} it is shown that the evolution equations determined by $\ell: (J^1P)/G \rightarrow \Lambda^{n+1}X$ and $l: (T \Gamma(P_M))/G \rightarrow \mathbb{R}$ are equivalent. As before, the reconstruction of solutions from the covariantly reduced system necessitates the flatness of the solution connection whereas the reconstruction of the dynamically reduced system does not need any additional hypotheses.
This difference is only apparent since, in the dynamic approach, the configuration manifold $\Gamma(P_M)/G$ is interpreted as the space of flat connections  and hence the compatibility condition appearing in covariant reduction is built into dynamic reduction.  
\medskip

The paper is structured in the following way. It begins with a presentation of the covariant Euler-Poincar\'e reduction in Section \ref{sec:covariant_EP}. All formulas are explicitly computed in the case of a trivial bundle over spacetime. In Section \ref{spin_systems} the affine Euler-Poincar\'e reduction in the formulation useful to the theory of spin systems is recalled; in particular the bundle is trivial. The passage from covariant to dynamical reduction in the case of trivial principal bundles and the equivalence of the reduced equations is presented in Section \ref{sec:trivial_bundles}. Affine Euler-Poincar\'e reduction for non-trivial principal bundles is formulated in Section \ref{sec:affine_EP_general}. Section \ref{sec:covariant_to_dynamic_general} contains the main result of the paper: the passage from covariant to dynamic reduction for general principal bundles and the equivalence of the two reduced systems.

\section{Covariant Euler-Poincar\'e reduction}
\label{sec:covariant_EP}

In this section we present all the background material for covariant reduction of principal bundle field theories. We begin by recalling the general covariant Lagrangian reduction procedure and work out in detail all formulas for trivial bundles which are needed in later sections.

\subsection{The general case}

Let $X$ be a $(n+1)$ dimensional manifold and let $\pi:P\rightarrow X$ be a \textit{right\/} principal bundle over $X$, with structure group $G$. For $U\subset X$ an open subset, we will denote by $\pi_U:P_U\rightarrow U$ the restricted principal bundle over $U$.

The first jet bundle is the affine bundle $J^1P\rightarrow P$ over $P$ whose fiber at $p$ is
\[
J^1_pP=\{\gamma_p\in L(T_xX,T_pP)\mid T_p\pi\!\cdot\!\gamma_p=id_{T_xX}\},
\]
where $x=\pi(p)$ and $L(T_xX,T_pP)$ denotes the linear maps $\gamma_p:T_xX\rightarrow T_pP$. Given a local section $s:U\subset X\rightarrow P_U$, the first jet extension of $s$ is the section $j^1s$ of the fiber bundle $J^1P_U\rightarrow U\subset X$ defined by $j^1s(m):=T_ms$. A global section exists if and only if the bundle is trivial. The structure group $G$ acts naturally on $J^1P$, the action being given by
\[
\gamma_p\mapsto T\Phi_g\!\cdot\!\gamma_p : = T\Phi_g \circ \gamma_p.
\]
The resulting quotient space $J^1P/G$ is an affine bundle over $X$, whose sections can be identified with principal connections on $P$.

A Lagrangian density is a smooth bundle map $\mathcal{L}:J^1P\rightarrow\Lambda^{n+1}X$ over $X$, where $n+1=\operatorname{dim} X$. It is said to be $G$ invariant if
\[
\mathcal{L}(T\Phi_g\!\cdot\!\gamma_p)=\mathcal{L}(\gamma_p),
\]
for all $g\in G$. In this case, $\mathcal{L}$ induces a reduced Lagrangian density $\ell:J^1P/G\rightarrow\Lambda^{n+1}X$. For simplicity, we suppose that $X$ is orientable and we fix a volume form $\mu$ on $X$. In this case we can write $\mathcal{L}=\bar{\mathcal{L}}\mu$, and $\ell=\bar{\ell}\mu$, where
\[
\bar{\mathcal{L}}:J^1P\rightarrow\mathbb{R},\quad \bar{\ell}:J^1P/G\rightarrow\mathbb{R}.
\]

Given a principal connection $\mathcal{A}$ on $P$ and denoting by $VP$ the vertical subbundle of $TP$, we have an affine bundle isomorphism over $P$ defined by
\begin{equation}\label{F_A_def}\mathcal{F}_\mathcal{A}:J^1P\rightarrow L(TX,VP),\quad\mathcal{F}_\mathcal{A}(\gamma_p):=\gamma_p-\operatorname{Hor}_p^\mathcal{A},
\end{equation}
where $L(TX,VP)\rightarrow P$ denotes the vector bundle whose fiber at $p$ is $L(T_x,V_pP)$ and $\operatorname{Hor}_p^\mathcal{A}$ denotes the horizontal lift with respect to $\mathcal{A}$. Note that for $\delta_p\in L(T_xX,V_pP)$ we have
\[
\mathcal{F}_\mathcal{A}(\gamma_p+\delta_p)=\mathcal{F}_\mathcal{A}(\gamma_p)+\delta_p.
\]

The map $\mathcal{F}_\mathcal{A}$ drops to the quotient spaces and gives an affine bundle isomorphism over $X$
\[
\Psi_\mathcal{A}:J^1P/G\rightarrow L(TX,VP/G), \quad \Psi_\mathcal{A}([\gamma_p]):=\left[\gamma_p-\operatorname{Hor}_p^\mathcal{A}\right].
\]
Note that for $[\delta_p]\in L(TX,VP/G)$, we have
\[
\Psi_\mathcal{A}([\gamma_p+\delta_p])=\Psi_\mathcal{A}([\gamma_p])+[\delta_p].
\]
We denote by $\operatorname{Ad}P:=(P\times\mathfrak{g})/G$ the adjoint bundle and by $[p,\xi]_G$ an element in the fiber $(\operatorname{Ad}P)_x, x=\pi(p)$. Recall that there is a vector bundle isomorphism
\[
\sigma:\operatorname{Ad}P\rightarrow VP/G
\]
over $X$, given by $\sigma_x\left([p,\xi]_G\right)=[\xi_P(p)]$, where 
\begin{equation}\label{infinitesimal}
\xi_P(p)=\left.\frac{d}{dt}\right|_{t=0}\Phi_{\operatorname{exp}(t\xi)}(p)\in V_pP
\end{equation}
is the infinitesimal generator. The inverse of $\sigma$ is given by $\sigma_x^{-1}[v_p]=[p,\mathcal{A}(v_p)]_G$. As a consequence, we can see $\Psi_\mathcal{A}$ as a bundle map over $X$ with values in $L(TX,\operatorname{Ad}P)$, given by
\[
\Psi_\mathcal{A}:J^1P/G\rightarrow L(TX,\operatorname{Ad}P),\quad [\gamma_p]\mapsto [p,\mathcal{A}\!\cdot\!\gamma_p]_G,
\]
where $L(TX,\operatorname{Ad}P)\rightarrow X$ denotes the vector bundle whose fiber at $x$ is $L(T_xX,\operatorname{Ad}P_x)$ and $(\mathcal{A} \!\cdot\! \gamma_p)(u_p) : = \mathcal{A}(p) \left(\gamma_p(u_p) \right)$ for any $u _p\in T _pP $.

The reduced Lagrangian density on $L(TX,\operatorname{Ad}P)$ is defined by $\bar\ell^\mathcal{A}:=\bar\ell\circ\Psi_\mathcal{A}^{-1}$.

\paragraph{Remarks.}
$(1)$ The isomorphism $\Psi_\mathcal{A}$ is the analog, for $J^1P/G$, of the connection dependent vector bundle isomorphism
\begin{equation}\label{identification}
TP/G\rightarrow TX\oplus_X\operatorname{Ad}P,\quad [u_p]\rightarrow \left(T_p\pi(u_p),[p,\mathcal{A}(u_p)]_G\right).
\end{equation}
$(2)$ Note that $J^1P/G$ is only an affine bundle. However the choice of a connection $\mathcal{A}$ allows one to endow $J^1P/G$ with the structure of a vector bundle, by pulling back the vector bundle structure of $L(TX,\operatorname{Ad}P)$. The zero element in the fiber at $x$ is $\left[\operatorname{Hor}_p^\mathcal{A}\right]$.\\
$(3)$ The fact that the affine bundle $J^1P/G$ has the vector bundle $L(TX,\operatorname{Ad}P)$ as underlying linear space reflects the fact that the sections of $J^1P/G$ are principal connections and that the affine space $\mathcal{C}onn(P)$ of all principal connections on $P$ has $\Omega^1(X,\operatorname{Ad}P)=\Gamma(L(TX,\operatorname{Ad}P))$ as underlying vector space. Given a section $\sigma$ of $J^1P/G$, the associated principal connection on $P \rightarrow  X $ is denoted by $\mathcal{A}^\sigma$ and is determined by the condition
\begin{equation}\label{condition_connection}
\left[\operatorname{Hor}^{\mathcal{A}^\sigma}_p(v_x)\right]=\sigma(x)(v_x),
\end{equation}
for all $p\in P$ and all $v_x\in T_xX, x=\pi(p)$.\\
$(4)$ Note that if $[\gamma_p]\in (J^1P/G)_x$, $x=\pi(p)$, then $[p,\mathcal{A}\!\cdot\!\gamma_p]_G\in L(T_xX,\operatorname{Ad}P_x)$. Indeed, applying $[p,\mathcal{A}\!\cdot\!\gamma_p]_G$ to a vector $v_x\in T_xX$, we get the vector
\[
[p,\mathcal{A}(\gamma_p(v_x))]_G\in \operatorname{Ad}P_x.
\]

Given a local section $s:U\subset X\rightarrow P_U$, its \textit{reduced first jet extension} is the local section $\sigma:U\subset X\rightarrow J^1P_U/G$ defined by
\[
\sigma(x):=[j^1s(x)].
\]
We will also define its \textit{reduced first jet extension associated to the connection} $\mathcal{A}$ given by the local section $\sigma^\mathcal{A}:U\subset X\rightarrow L\left(TU,\operatorname{Ad}P_U\right)$ by
\[
\sigma^\mathcal{A}(x):=\Psi_\mathcal{A}([j^1s(x)])=[s(x),\mathcal{A}\!\cdot\!T_xs]_G.
\]

Note that a section of the vector bundle $L\left(TU,\operatorname{Ad}P_U\right)\rightarrow U\subset X$ can be interpreted as a one-form on $U\subset X$ taking values in the vector bundle $\operatorname{Ad}P_U$, that is,
\[
\Gamma\left(L\left(TU,\operatorname{Ad}P_U\right)\right)=\Omega^1\left(U,\operatorname{Ad}P_U\right).
\]
For example, we have $\sigma^\mathcal{A}\in \Omega^1(U,\operatorname{Ad}P_U)$.
\medskip

We now state the covariant Euler-Poincar\'e reduction theorem, see \cite{CaRaSh2000}.

\begin{theorem} Let $\pi:P\rightarrow X$ be a right principal $G$-fiber bundle over a manifold $X$ with volume form $\mu$ and let $\mathcal{L}:J^1P\rightarrow\Lambda^{n+1}X$ be a $G$ invariant Lagrangian density. Let $\ell: J^1P/G\rightarrow \Lambda^{n+1}X$ be the reduced Lagrangian density associated to $L$. For a local section $s:U\subset X\rightarrow P_U$, let $\sigma:U\rightarrow J^1P_U/G$ be the reduced first jet extension of $s$ and let $\sigma^\mathcal{A}:U\rightarrow L\left(TU,\operatorname{Ad}P_U\right)$ be defined as before, where $\mathcal{A}$ is a principal connection on the bundle $P_U\rightarrow U$. Then the following are equivalent:
\begin{itemize}
\item The variational principle
\[
\delta\int_U\mathcal{L}(j^1s)=0
\]
holds, for vertical variations along $s$ with compact support.
\item The local section $s$ satisfies the \textbf{covariant Euler-Lagrange equations} for $\mathcal{L}$.
\item The variational principle
\[
\delta\int_U\ell(\sigma)=0
\]
holds, using variations of the form
\[
\delta\sigma=\nabla^\mathcal{A}\eta-[\sigma^\mathcal{A},\eta],
\]
where $\eta:U\subset X\rightarrow \operatorname{Ad}P_U$ is a section with compact support, and
\[
\nabla^\mathcal{A}:\Gamma(\operatorname{Ad}P_U)\rightarrow \Omega^1(U,\operatorname{Ad}P_U)
\]
denotes the affine connection induced by the principal connection $\mathcal{A}$ on $P_U$.
\item The \textbf{covariant Euler-Poincar\'e equations} hold:
\[
\operatorname{div}^\mathcal{A}\frac{\delta \bar{\ell}}{\delta\sigma}=-\operatorname{Tr}\left(\operatorname{ad}^*_{\sigma^\mathcal{A}}\frac{\delta \bar{\ell}}{\delta\sigma}\right),
\]
where $\frac{\delta \bar{\ell}}{\delta\sigma}$ is the section of the vector bundle $L\left(TU,\operatorname{Ad}P_U\right)^*=L\left(T^*U,\operatorname{Ad}P^*_U
\right)$ defined by
\[
\frac{\delta \bar{\ell}}{\delta\sigma}(\zeta_x):=\left.\frac{d}{dt}\right|_{t=0}\bar{\ell}(\sigma(x)+t\zeta_x),
\]
for $\zeta_x\in L(T_xX,\operatorname{Ad}P_x), x\in U$, and where
\[
\operatorname{div}^\mathcal{A}:\Gamma(L(T^*U,\operatorname{Ad}P^*_U))=\mathfrak{X}(U,\operatorname{Ad}P^*_U)\rightarrow\Gamma(\operatorname{Ad}P^*_U).
\]
denotes the covariant divergence associated to the connection $\mathcal{A}$, defined by the condition
\[
\operatorname{div}(w\!\cdot\!\eta)=\left(\operatorname{div}^\mathcal{A}w\right)\!\cdot\!\eta+w\!\cdot\!\nabla^\mathcal{A}\eta,
\]
for all $w\in\mathfrak{X}(U,\operatorname{Ad}P^*)$ and $\eta\in \Gamma(\operatorname{Ad}P)$.
\end{itemize}
\end{theorem}

\paragraph{Remarks.} $(1)$ One can also write the covariant Euler-Poincar\'e equations in terms of $\bar\ell^\mathcal{A}$ on $L(TU,\operatorname{Ad}P)$. Using the equality
\[
\frac{\delta\bar\ell^\mathcal{A}}{\delta\sigma^\mathcal{A}}=\frac{\delta\bar\ell}{\delta\sigma},
\]
the covariant Euler-Poincar\'e equations read
\begin{align}\label{Cov_EP_A}
\operatorname{div}^\mathcal{A}\frac{\delta\bar\ell^\mathcal{A}}{\delta\sigma^\mathcal{A}}=-\operatorname{Tr}\left(\operatorname{ad}^*_{\sigma^\mathcal{A}}\frac{\delta\bar\ell^\mathcal{A}}{\delta\sigma^\mathcal{A}}\right).
\end{align}
$(2)$ Not any solution of the covariant Euler-Poincar\'e equations comes from a solution
of the original covariant Euler-Lagrange equations of $\mathcal{L}$. An extra equation, a \textit{compatibility condition}, must be imposed. This condition simply reads
\[
\mathbf{d}^{\mathcal{A}^\sigma}\mathcal{A}^\sigma = 0;
\]
that is, the principal connection $\mathcal{A}^\sigma$ associated to the critical section $\sigma$ must be flat.

\subsection{The case of a trivial principal bundle}

Let us work out the previous theory in the case of a \textit{right\/} trivial principal  $G$-bundle
\[
\pi:P=X\times G\rightarrow X,\quad \pi(p)=x,
\]
where $p=(x,g)$. The tangent map to the projection reads simply
\[
T\pi:TX\times TG\rightarrow TX,\quad T\pi(u_x,\xi_g)=u_x.
\]
The first jet bundle $J^1P\rightarrow P$ can be identified with the vector bundle $L(TX,TG)\rightarrow P$. More precisely, we have $J^1P_{(x,g)}\simeq L(T_xX,T_gG)$. Indeed, any $\gamma_{(x,g)}\in L(T_xX,T_xX\times T_gG)$ reads
\[
\gamma_{(x,g)}(u_x)=\left(a_x(u_x),b_{(x,g)}(u_x)\right),
\]
where $a_x\in L(T_xX,T_xX)$ and $b_{(x,g)}\in L(T_xX,T_gG)$. Thus, the condition $T\pi(\gamma_{(x,g)}(u_x))=u_x$ reads
\[
a_x=id_{T_xX}.
\]
As a consequence, any $\gamma_{(x,g)}\in J^1P_{(x,g)}$ is of the form
\[
\gamma_{(x,g)}(u_x)=(u_x,b_{(x,g)}(u_x)),\quad b_{(x,g)}\in L(T_xX,T_gG),
\]
that is, we have $\gamma_{(x,g)}=id_{T_xX}\times b_{(x,g)}$, and we can identify $\gamma_{(x,g)}$ with $b_{(x,g)}$. This proves that
\[
J^1P_{(x,g)}\simeq L(T_xX,T_gG).
\]

Concerning the reduction of the tangent bundle, the tangent lifted action of $G$ on $TP$ reads $(u_x,\xi_h)\mapsto (u_x,TR_g\xi_h)$. Thus the class $[u_x,\xi_g]$ can be identified with the element $(u_x,TR_{g^{-1}}\xi_g)$. More precisely, we have the vector bundle isomorphism
\[
TP/G\rightarrow TX\times\mathfrak{g},\quad [u_x,\xi_g]\rightarrow (u_x,TR_{g^{-1}}\xi_g).
\]
This isomorphism corresponds to the choice of the standard connection $\mathcal{A}(x,g)(u_x,\xi_g)=TL_{g^{-1}}\xi_g$ in the identification \eqref{identification}, that is, the horizontal space $H_{(x,g)}P= T_x X \times \{0\}$. In the case of a general connection $\mathcal{A}$ on $P$, the vector bundle isomorphism \eqref{identification} reads
\[
TP/G\rightarrow TX\times\mathfrak{g},\quad [u_x,\xi_g]\rightarrow \left(u_x,\overline{\mathcal{A}}(x)(u_x)+TR_{g^{-1}}\xi_g\right),
\]
where $\overline{\mathcal{A}}\in\Omega^1(X,\mathfrak{g})$, denotes the local expression of $\mathcal{A}$, that is,
\[
\mathcal{A}(x,g)(u_x,\xi_g)=\operatorname{Ad}_{g^{-1}}\left(\overline{\mathcal{A}}(x)(u_x)+TR_{g^{-1}}\xi_g\right).
\]

Concerning the reduction of the first jet bundle, the action of $G$ on $\gamma_{(x,g)}\in J^1P\simeq L(TX,TG)$ reads 
\[
b_{(x,g)}\rightarrow TR_h\!\cdot\! b_{(x,g)},
\]
and we can identify the class $[\gamma_{(x,g)}]$ with the element $(id_{T_xX},TR_g^{-1}\!\cdot\! b_{(x,g)})$ and we have the bundle isomorphism
\[
J^1P/G\rightarrow L(TX,\mathfrak{g}),\quad [\gamma_{(x,g)}]\mapsto TR_g^{-1}\!\cdot\! b_{(x,g)}.
\]
This isomorphism corresponds to the choice of the standard connection for $\Psi_\mathcal{A}$. In general, the map $\Psi_\mathcal{A}$ reads
\[
J^1P/G\rightarrow L(TX,\mathfrak{g}),\quad [\gamma_{(x,g)}]\rightarrow \overline{\mathcal{A}}(x)+TR_{g^{-1}}\!\cdot\! b_{(x,g)},
\]
since $\gamma_{(x,g)}=id_{T_xX}\times b_{(x,g)}$. Indeed,
\begin{align*}
\Psi_\mathcal{A}\left(\gamma_{(x,g)}\right)&=\left[(x,g),\mathcal{A}(x,g)\!\cdot\!\left(id_{T_xX},b_{(x,g)}\right)\right]_G\\
&=\left[(x,g),\operatorname{Ad}_{g^{-1}}\left(\overline{\mathcal{A}}(x)+TR_{g^{-1}}\!\cdot\!b_{(x,g)}\right)\right]_G\\
&=\left[(x,e),\overline{\mathcal{A}}(x)+TR_{g^{-1}}\!\cdot\!b_{(x,g)}\right]_G.
\end{align*}
Thus, using the identification of $\Omega^1(U,\operatorname{Ad}P_U)$ with $\Omega^1(U,\mathfrak{g})$ valid in the trivial case, we can identify $\Psi_\mathcal{A}(\gamma_{(x,g)})$ with
\[
\overline{\mathcal{A}}(x)+TR_{g^{-1}}\!\cdot\! b_{(x,g)}.
\]

A local section $s:U\subset X\rightarrow P$ reads simply $s(x)=(x,\bar s(x))$ where $\bar s\in\mathcal{F}(U,G)$, and its first jet extension is $j^1s(x)(u_x)=T_xs(u_x)=(u_x,T_x\bar s(u_x))\in J^1P_{(x,\bar s(x))}$ which can be identified with $T_x\bar s(u_x)=j^1\bar s(x)\in L(T_xX,T_{\bar s(x)}G)$.

The reduced first jet extension $\sigma:=[j^1s]$ reads
\[
\sigma(x)=[id_{T_xX},T_x\bar s],
\]
and this class can be identified with $\bar\sigma(x)=TR_{\bar s(x)}^{-1}\!\cdot\!T_x\bar s$. As before, this corresponds to the choice of the standard connection for $\Psi_\mathcal{A}$. Note that, by \eqref{condition_connection}, the principal connection associated to $\sigma$ is
\[
\overline{\mathcal{A}^\sigma}(x)=-TR_{\bar s(x)^{-1}}\!\cdot\!T_x\bar s=-\bar\sigma(x).
\]
For a general connection $\mathcal{A}$, the reduced first jet extension $\sigma^\mathcal{A}\in \Omega^1(U,\operatorname{Ad}P_U)$ associated to $\mathcal{A}$ can be written
\begin{align*}
\sigma^\mathcal{A}(x):&=\Psi_\mathcal{A}([j^1s(x)])=\left[s(x),\mathcal{A}\!\cdot\!T_xs\right]_G\\
&=\left[(x,\bar s(x)),\mathcal{A}\!\cdot\!\left(id_{T_xX},T_x\bar s\right)\right]_G\\
&=\left[(x,\bar s(x)),\operatorname{Ad}_{\bar s(x)^{-1}}\left(\overline{\mathcal{A}}(x)+TR_{\bar s(x)^{-1}}\!\cdot\!T_x\bar s\right)\right]_G\\
&=\left[(x,e),\overline{\mathcal{A}}(x)+TR_{\bar s(x)^{-1}}\!\cdot\!T_x\bar s\right]_G\\
&=:\left[(x,e),\bar \sigma^\mathcal{A}(x)\right]_G.
\end{align*}
Thus, using the identification $\Omega^1(U,\operatorname{Ad}P_U)\simeq \Omega^1(U,\mathfrak{g})$ valid in the case of a trivial bundle, we can identify $\sigma^\mathcal{A}$ with $\bar\sigma^\mathcal{A}$ given by
\[
\bar \sigma^\mathcal{A}(x)=\overline{\mathcal{A}}(x)+TR_{\bar s(x)^{-1}}\!\cdot\!T_x\bar s=\overline{\mathcal{A}}(x)+\bar\sigma(x)\in L(TX,\mathfrak{g}).
\]

As a consequence, in the trivial case, the covariant Euler-Poincar\'e equations
\[
\operatorname{div}^\mathcal{A}\frac{\delta \bar\ell}{\delta \sigma}=-\operatorname{ad}^*_{\sigma^\mathcal{A}}\frac{\delta \bar\ell}{\delta \sigma},
\]
read
\[
\operatorname{div}^{\overline{\mathcal{A}}}\frac{\delta \bar\ell}{\delta \bar\sigma}=-\operatorname{ad}^*_{\overline{\mathcal{A}}+\bar\sigma}\frac{\delta \bar\ell}{\delta \bar\sigma}
\]
or, equivalently,
\[
\operatorname{div}\frac{\delta \bar\ell}{\delta \bar\sigma}=-\operatorname{ad}^*_{\bar\sigma}\frac{\delta \bar\ell}{\delta \bar\sigma}\quad\text{or}\quad \operatorname{div}^{-\bar\sigma}\frac{\delta \bar\ell}{\delta\bar\sigma}=0.
\]
Similarly, the constrained variations
\[
\delta\bar\sigma=\mathbf{d}^{\overline{\mathcal{A}}}\eta-[\bar\sigma^\mathcal{A},\eta]
\]
become
\[
\delta\bar\sigma=\mathbf{d}\eta-[\bar\sigma,\eta].
\]
In the case of a trivial principal bundle, the covariant Euler-Poincar\'e reduction theorem can be stated as follows.

\begin{theorem} Let $\pi:P=X\times G\rightarrow X$ be a trivial right principal bundle, with structure group $G$, over a manifold $X$ endowed with a volume form $\mu$. Let $\mathcal{L}:J^1P\rightarrow\Lambda^{n+1}X$ be a $G$ invariant Lagrangian density. Let $\ell: J^1P/G\simeq L(TX,\mathfrak{g})\rightarrow \Lambda^{n+1}X$ be the reduced Lagrangian density associated to $\mathcal{L}$. For a local section $\bar s:U\subset X\rightarrow G$, let $\bar\sigma, \bar\sigma^\mathcal{A} :U\rightarrow L(TU,\mathfrak{g})$ be defined by
\[
\bar\sigma(x)=TR_{\bar s(x)^{-1}}\!\cdot\! T_x\bar s\quad\text{and}\quad\bar\sigma^\mathcal{A}(x)=\overline{\mathcal{A}}(x)+TR_{\bar s(x)^{-1}}\!\cdot\! T_x\bar s\in L(TU,\mathfrak{g}),
\]
where $\mathcal{A}$ is a principal connection on the bundle $P_U=U\times G\rightarrow U$. Then the following are equivalent:
\begin{itemize}
\item The variational principle
\[
\delta\int_U\mathcal{L}(j^1s)=0
\]
holds, for vertical variations along $s$ with compact support.
\item The local section $s$ satisfies the \textbf{covariant Euler-Lagrange equations} for $\mathcal{L}$.
\item The variational principle
\[
\delta\int_U\ell(\bar\sigma(x))=0
\]
holds, using variations of the form
\[
\delta\bar\sigma=\mathbf{d}^{\overline{\mathcal{A}}}\eta-\left[\bar\sigma^\mathcal{A},\eta\right] = \mathbf{d}\eta -\left[\bar{ \sigma}, \eta \right],
\]
where $\eta:U\subset X\rightarrow \mathfrak{g}$ has compact support, and
\[
\mathbf{d}^{\overline{\mathcal{A}}}:\mathcal{F}(U,\mathfrak{g})\rightarrow\Omega^1(U,\mathfrak{g}),\quad\mathbf{d}^{\overline{\mathcal{A}}}\eta=\mathbf{d}f+\left[\,\overline{\mathcal{A}},f\right]
\]
is the covariant differential on trivial principal bundles.
\item The \textbf{covariant Euler-Poincar\'e equations} hold:
\begin{equation}
\label{covariant_EP_trivial}
\operatorname{div}^{\overline{\mathcal{A}}}\frac{\delta \bar{\ell}}{\delta\bar\sigma}=-\operatorname{Tr}\left(\operatorname{ad}^*_{\bar\sigma^\mathcal{A}}\frac{\delta \bar{\ell}}{\delta\bar\sigma}\right),
\end{equation}
where $\frac{\delta \bar{\ell}}{\delta\bar\sigma}$ is the section of the vector bundle $L(TU,\mathfrak{g})^*=L(T^*U,\mathfrak{g}^*)$ defined by
\[
\frac{\delta \bar{\ell}}{\delta\bar\sigma}(\zeta_x):=\left.\frac{d}{dt}\right|_{t=0}\bar{\ell}(\bar\sigma(x)+t\zeta_x),
\]
for $\zeta_x\in L(T_xX,\mathfrak{g}), x\in U$, and where 
\[
\operatorname{div}^{\overline{\mathcal{A}}}:\Gamma(L(T^*U,\mathfrak{g}^*))=\mathfrak{X}(U,\mathfrak{g}^*)\rightarrow\mathcal{F}(U,\mathfrak{g}^*),\quad\operatorname{div}^{\overline{\mathcal{A}}}w=\operatorname{div}w-\operatorname{Tr}\left(\operatorname{ad}^*_{\overline{\mathcal{A}}}w\right)
\]
is the covariant divergence (for a trivial principal bundle) associated to the connection $\mathcal{A}$.
\end{itemize}
\end{theorem}

\section{Affine Euler-Poincar\'e for spin systems}\label{spin_systems}

We now recall from \cite{GBRa2008b} the process of affine Euler-Poincar\'e reduction as it applies to spin systems. This general procedure explains the geometric structure of many complex fluid models. The key idea of this method is to introduce new advection equations containing affine terms.
\medskip

Given a manifold $M$ and a Lie group $G$, we can form the group $\mathcal{F}(M,G)$ of $G$-valued maps on $M$.

\begin{itemize}
\item Assume that we have a Lagrangian
\[
L:T\mathcal{F}(M,G)\times\Omega^1(M,\mathfrak{g})\rightarrow\mathbb{R}
\]
which is \textit{right\/} invariant under the affine action of $\Lambda\in\mathcal{F}(M,G)$ given by
\[
(\chi,\dot\chi,\gamma)\mapsto (\chi\Lambda,\dot\chi\Lambda,\theta_{\Lambda}(\gamma)),\quad \theta_{\Lambda}(\gamma):=\Lambda^{-1}\gamma\Lambda+\Lambda^{-1}\mathbf{d}\Lambda.
\]
\item For a fixed $\gamma_0\in\Omega^1(M,\mathfrak{g})$, we define the Lagrangian $L_{\gamma_0}:T\mathcal{F}(M,G)\rightarrow\mathbb{R}$ by $L_{\gamma_0}(\chi,\dot\chi):=L(\chi,\dot\chi,\gamma_0)$. Then $L_{\gamma_0}$ is right invariant under the lift to $T\mathcal{F}(M,G)$ of the right action
of $\mathcal{F}(M,G)_{\gamma_0}$ on $\mathcal{F}(M,G)$, where $\mathcal{F}(M,G)_{\gamma_0}$ denotes the isotropy group of $\gamma_0$ with respect to the affine action $\theta$.
\item Right invariance of $L$ permits us to define the reduced Lagrangian $l=l(\nu,\gamma):\mathcal{F}(M,\mathfrak{g})\times\Omega^1(M,\mathfrak{g})\rightarrow\mathbb{R}$.
\item For a curve $\chi(t)\in \mathcal{F}(M,G)$, let $\nu(t):=\dot\chi(t)\chi(t)^{-1}\in\mathcal{F}(M,\mathfrak{g})$ and define the curve $\gamma(t)$ as the unique solution of the following affine differential equation with time dependent coefficients
\[
\dot\gamma+\mathbf{d}^\gamma\nu=0,
\]
where $\mathbf{d}^\gamma\nu=\mathbf{d}\nu+\operatorname{ad}_\gamma\nu$ and with initial condition $\gamma_0$. The solution can be written as
\[
\gamma(t)=\theta_{\chi(t)^{-1}}\gamma_0=\chi(t)\gamma_0\chi(t)^{-1}+\chi(t)\mathbf{d}\chi(t)^{-1}.
\]
\end{itemize}

\begin{theorem}\label{AEP_thm} With the preceding notations, the following are equivalent:
\begin{itemize}
\item With $\gamma_0$ held fixed, Hamilton's variational principle
\[
\delta\int_{t_1}^{t_2}L_{\gamma_0}(\chi(t),\dot\chi(t))dt=0,
\]
holds, for variations $\delta\chi(t)$ of $\chi(t)$ vanishing at the endpoints.
\item $\chi(t)$ satisfies the Euler-Lagrange equations for $L_{\gamma_0}$ on $\mathcal{F}(M,G)$.
\item The constrained variational principle
\[
\delta\int_{t_1}^{t_2}l(\nu(t),\gamma(t))dt=0,
\]
holds on $\mathcal{F}(M,\mathfrak{g})\times \Omega^1(M,\mathfrak{g})$, upon using variations of the form
\[
\delta\nu=\dot\zeta-[\nu,\zeta],\quad\delta\gamma=-\mathbf{d}^\gamma\zeta,
\]
where $\zeta(t)\in\mathcal{F}(M,\mathfrak{g})$ vanishes at the endpoints.
\item The \textbf{affine Euler-Poincar\'e equations} hold on $\mathcal{F}(M,\mathfrak{g})\times\Omega^1(M,\mathfrak{g})$:
\[
\frac{\partial }{\partial t}\frac{\delta l}{\delta\nu}=-\operatorname{ad}^*_\nu\frac{\delta l}{\delta\nu}+\operatorname{div}^\gamma\frac{\delta l}{\delta\gamma}.
\]
\end{itemize}
\end{theorem}

Note that this reduction process generalizes easily to the case of a time dependent Lagrangian.

In order to formulate the affine Euler-Poincar\'e equations above, one has to choose spaces in (weak) nondegenerate duality with the spaces  $\mathcal{F}(M,\mathfrak{g})$ and $\Omega^1(M,\mathfrak{g})$, relative to a pairing $\langle\,,\rangle$. The associated functional derivative is defined by
\[
\left\langle\frac{\delta l}{\delta\nu},\zeta\right\rangle=\left.\frac{d}{dt}\right|_{t=0}l(\nu+t\zeta).
\]
It will be convenient to choose as dual spaces $\mathcal{F}(M,\mathfrak{g}^*)$ and $\mathfrak{X}(M,\mathfrak{g}^*)$, respectively, where the duality pairing is given by contraction followed by integration over $M$, with respect to a fixed volume form on $M$.

\section{Covariant to dynamical reduction for trivial bundles}
\label{sec:trivial_bundles}

In this section we work exclusively with trivial principal bundles. We introduce a new dynamical Lagrangian associated to a symmetric field theoretic Lagrangian density and show that it is gauge group invariant. Covariant and dynamic reduction is performed and it is shown that the resulting reduced equations are equivalent.

\medskip

Given a manifold $M$ we consider the manifold $X=\mathbb{R}\times M$ which will plays the role of spacetime. For simplicity we suppose that $M$ is orientable, with volume form $\mu_M$, and we endow $X$ with the volume form $\mu=dt\wedge\mu_M$. Given a Lie group $G$, we consider the trivial principal bundles
\[
\pi_M:P_M:=M\times G\rightarrow M\quad\text{and}\quad \pi: P_X:=X\times G\rightarrow X.
\]
Recall the we have the vector bundle isomorphisms
\[
J^1P_M\rightarrow L(TM,TG)\quad\text{and}\quad J^1P_X\rightarrow L(TX,TG)=L(T\mathbb{R},TG)\times_{P_X} L(TM,TG)
\]
over $P_M$ and $P_X$ respectively. Note that every element $b_{(x,g)}\in J^1P\simeq L(TX,TG)$ reads
\begin{equation}\label{splitting}
b_{(x,g)}=\left(b_{(t,g)},b_{(m,g)}\right),
\end{equation}
where $b_{(t,g)}\in L(T_t\mathbb{R},T_gG)$ and $b_{(m,g)}\in L(T_mM,T_gG)$.

We consider a $G$ invariant Lagrangian density $\mathcal{L}:J^1P_X\rightarrow\Lambda^{n+1}X$. Using the decomposition \eqref{splitting}, we can write
\[
\mathcal{L}(b_{(x,g)})=\mathcal{L}\left(b_{(t,g)},b_{(m,g)}\right).
\]

We are now ready to define the main object of this paper.

\begin{definition} Let $\mathcal{L}:J^1P_X\rightarrow\Lambda^{n+1}X$ be a $G$-invariant Lagrangian density. The \textbf{instantaneous Lagrangian} is defined by
\[
L^{\mathcal{L}}=L^{\mathcal{L}}(t,\chi,\dot\chi,\gamma):I\times T\mathcal{F}(M,G)\times\Omega^1(M,\mathfrak{g})\rightarrow\mathbb{R},
\]
\begin{equation}\label{definition_L}
L^{\mathcal{L}}(t,\chi,\dot\chi,\gamma):=\int_M\overline{\mathcal{L}}(t,\dot\chi(m),\mathbf{d}\chi(m)-\chi(m)\gamma(m))\mu_M.
\end{equation}
\end{definition}

The Lagrangian $L^{\mathcal{L}}$ has the remarkable property to be invariant under the affine action of $\mathcal{F}(M,G)$. This is stated in the following theorem.

\begin{theorem} Consider a $G$ invariant Lagrangian density $\mathcal{L}:J^1P\rightarrow \Lambda^{n+1}X$ and its associated Lagrangian $L^{\mathcal{L}}$ defined in \eqref{definition_L}. Then for all $\Lambda\in\mathcal{F}(M,G)$, we have
\[
L^{\mathcal{L}}(t,\chi\Lambda,\dot\chi\Lambda,\Lambda^{-1}\gamma\Lambda+\Lambda^{-1}\mathbf{d}\Lambda)=L^{\mathcal{L}}(t,\chi,\dot\chi,\gamma).
\]
\end{theorem}
\textbf{Proof.} We have
\begin{align*}
L^{\mathcal{L}}(\chi\Lambda,\dot\chi\Lambda,\Lambda^{-1}\gamma\Lambda+\Lambda^{-1}\mathbf{d}\Lambda)&=\int_M\mathcal{L}(t,\dot\chi\Lambda,\mathbf{d}(\chi\Lambda)-\chi\Lambda(\Lambda^{-1}\gamma\Lambda+\Lambda^{-1}\mathbf{d}\Lambda))\\
&=\int_M\mathcal{L}(t,\dot\chi\Lambda,\mathbf{d}\chi\Lambda+\chi\mathbf{d}\Lambda-\chi\gamma\Lambda-\chi\mathbf{d}\Lambda)\\
&=\int_M\mathcal{L}(t,\dot\chi\Lambda,(\mathbf{d}\chi-\chi\gamma)\Lambda)=\int_M\mathcal{L}(t,\dot\chi,\mathbf{d}\chi-\chi\gamma)\\
&=L^{\mathcal{L}}(t,\chi,\dot\chi,\gamma).\qquad\blacksquare\\
\end{align*}

Recall that given an affine invariant, possibly time dependent, Lagrangian $L$ on $T\mathcal{F}(M,G)\times\Omega^1(M,\mathfrak{g})$ and a fixed $\gamma_0\in\Omega^1(M,\mathfrak{g})$ we can reduce the Euler-Lagrange equation for $L_{\gamma_0}$ and obtain the affine Euler-Poincar\'e equations
\begin{equation}\label{AEP}
\frac{\partial }{\partial t}\frac{\delta l}{\delta\nu}=-\operatorname{ad}^*_\nu\frac{\delta l}{\delta\nu}+\operatorname{div}^\gamma\frac{\delta l}{\delta\gamma},
\end{equation}
where $\nu=\dot\chi\chi^{-1}$ and $\gamma=\chi\gamma_0\chi^{-1}+\chi\mathbf{d}\chi^{-1}$.

On the other hand, since the Lagrangian density $\mathcal{L}$ on $J^1P$ is invariant, we can reduce the covariant Euler-Lagrange equations for $\mathcal{L}$ and obtain the covariant Euler-Poincar\'e equations for the reduced Lagrangian density $\ell$. Using that $X=\mathbb{R}\times M$, we obtain that the jet bundle $J^1P$ is isomorphic to the vector bundle $L(T\mathbb{R},TG)\times L(TM,TG)$. Thus, given a local section $\bar s=\bar s(t,m)\in\mathcal{F}(U,G)$ of $P$, where $U=I\times V\subset \mathbb{R}\times M=X$, its first jet extension reads
\[
T\bar{s}=(\dot{\bar{s}},\mathbf{d} \bar{s}).
\]
Similarly, the fiber  $(J^1P/G)_x$ of the reduced jet bundle is isomorphic to the vector space
\[
L(T_xX,\mathfrak{g}) =  L(T_t\mathbb{R},\mathfrak{g})\times L(T_mM,\mathfrak{g}),\quad x = (t, m).
\]
Thus, the reduced first jet bundle extension $\bar\sigma^\mathcal{A}\in\Omega^1(I\times V,\mathfrak{g})$ reads
\[
\bar\sigma^\mathcal{A}=(\bar\sigma^1,\bar\sigma^2),
\]
and we have $\bar\sigma^1_t:=\bar\sigma^1(t,\_\,)\in\mathcal{F}(M,\mathfrak{g})$ and $\bar\sigma^2_t:=\bar\sigma^2(t,\_\,)\in\Omega^1(M,\mathfrak{g})$.

Note that a principal connection $\mathcal{A}$ on $P_U$ reads
\[
\mathcal{A}(x,g)(u,u_m,\xi_g)=\operatorname{Ad}_{g^{-1}}\left(\overline{\mathcal{A}}^1(x)u+\overline{\mathcal{A}}^2(x)(u_m)+TR_{g^{-1}}\xi_g\right),\quad x=(t,m)\in U=I\times V.
\]
Thus, we obtain that $\bar\sigma^\mathcal{A}=(\bar\sigma^1,\bar\sigma^2)$ is given by
\[
\bar\sigma^1(x)=\overline{\mathcal{A}}^1(x)+TR_{\bar{s}(x)^{-1}}\dot{\bar{s}}(x),\quad \bar\sigma^2(x)=\overline{\mathcal{A}}^2(x)+TR_{\bar{s}(x)^{-1}}\mathbf{d}\bar{s}(x)
\]
Choosing the standard connection ($TL_g^{-1}\xi_g$) on $P$, the covariant Euler-Poincar\'e equations \eqref{covariant_EP_trivial} become in this case
\[
\frac{\partial}{\partial t}\frac{\delta\bar\ell}{\delta\bar\sigma^1}+\operatorname{div}\frac{\delta\bar\ell}{\delta\bar\sigma^2}=-\operatorname{ad}^*_{\bar\sigma^1}\frac{\delta\bar\ell}{\delta\bar\sigma^1}-\operatorname{Tr}\left(\operatorname{ad}^*_{\bar\sigma^2}\frac{\delta\bar\ell}{\delta\bar\sigma^2}\right),
\]
which can be rewritten as
\begin{equation}\label{cov_EP_3+1}
\frac{\partial}{\partial t}\frac{\delta\bar\ell}{\delta\bar\sigma^1}=-\operatorname{ad}^*_{\bar\sigma^1}\frac{\delta\bar\ell}{\delta\bar\sigma^1}+\operatorname{div}^{(-\bar\sigma_2)}\left(-\frac{\delta\bar\ell}{\delta \bar\sigma^2}\right),
\end{equation}
for
\[
\bar{\sigma}^1=TR_{\bar{s}^{-1}}\!\cdot\!\dot{\bar{s}}\quad\text{and}\quad \bar{\sigma}^2=TR_{\bar{s}^{-1}}\!\cdot\!\mathbf{d}\bar{s}.
\]

Note the remarkable fact that equations \eqref{AEP} and \eqref{cov_EP_3+1} are identical! Indeed, it suffices to set $\nu(t)(m)=\bar{\sigma}^1(x), \chi(t)(m)=\bar{s}(x), \gamma(t)(m)=-\bar{\sigma}^2(x)$, and $\gamma_0=0$. This fact is explained in the following theorem, which is the main result of this section.

Note that, in \eqref{AEP} and \eqref{cov_EP_3+1}, the functional derivatives of $\bar\ell$ and $l$ belong to the same space in spite of the fact that the Lagrangians $\bar\ell$ and $l$ are defined on different spaces. This is explained by the fact that the respective functional derivatives are defined differently. The functional derivatives of $\bar\ell$ are simply fiber derivatives whereas the functional derivatives of $l$ involve integration.

\begin{theorem}
\label{main_theorem_trivial_bundle}
 Consider a local section $\bar s=\bar s(x):U=I\times V\rightarrow G$ of the trivial principal bundle $X\times G\rightarrow X$, where $X=\mathbb{R}\times M$ and $x=(t,m)$, and its reduced first jet extension $\bar{\sigma}=(\bar{\sigma}^1,\bar{\sigma}^2) \in \Omega^1(I\times V,\mathfrak{g})$.

Define the curve $\chi(t)\in\mathcal{F}(V,G)$ by $\chi(t)(m):=\bar{s}(x)$, and the curves $\nu(t):=\dot{\chi}(t)\chi(t)^{-1}$ and $\gamma(t):=-\mathbf{d}\chi(t)\chi(t)^{-1}\in\Omega^1(V,\mathfrak{g})$. Thus we have
\[
\nu(t)(m)=\bar\sigma^1(x)\quad\text{and}\quad \gamma(t)(m)=-\bar\sigma^2(x).
\]

Consider a Lagrangian density
$\mathcal{L}:J^1P\rightarrow\Lambda^{n+1}X$ and define the associated time dependent Lagrangian $L^{\mathcal{L}}_{\gamma_0}:I\times T\mathcal{F}(V,G)\rightarrow\mathbb{R}$ by
\[
L_{\gamma_0}(t,\chi,\dot\chi):=\int_V\overline{\mathcal{L}}(t,\dot\chi,\mathbf{d}\chi-\chi\gamma_0)\mu_V,
\]
for all $\gamma_0\in\Omega^1(V,\mathfrak{g})$.

Then the corresponding reduced Lagrangians verify the relation
\[
l(t,\nu,\gamma)=\int_V \bar{\ell}(t,\nu,-\gamma)\mu_V
\]
and the following are equivalent:
\begin{itemize}
\item[{\bf (i)}]  Hamilton's variational principle
\[
\delta\int_{t_1}^{t_2}L_0(t,\chi(t),\dot\chi(t))dt=0,
\]
holds for variations $\delta\chi(t)$ of $\chi(t)$ vanishing at the endpoints.
\item[{\bf (ii)}] The curve $\chi(t)$ satisfies the Euler-Lagrange equations for $L_0$ on $\mathcal{F}(V,G)$.
\item[{\bf (iii)}] The constrained variational principle
\[
\delta\int_{t_1}^{t_2}l(t,\nu(t),\gamma(t))dt=0,
\]
holds on $\mathcal{F}(V,\mathfrak{g})\times\Omega^1(V,\mathfrak{g})$, upon using variations of the form
\[
\delta\nu=\dot\zeta-[\nu,\zeta],\quad\delta\gamma=-\mathbf{d}^\gamma\zeta,
\]
where $\zeta(t)\in\mathcal{F}(V,\mathfrak{g})$ vanishes at the endpoints.
\item[{\bf (iv)}] The \textbf{affine Euler-Poincar\'e equations} hold on $\mathcal{F}(V,\mathfrak{g})\times\Omega^1(V,\mathfrak{g})$:
\[
\frac{\partial }{\partial t}\frac{\delta l}{\delta\nu}=-\operatorname{ad}^*_\nu\frac{\delta l}{\delta\nu}+\operatorname{div}^\gamma\frac{\delta l}{\delta\gamma}.
\]
\item[{\bf (v)}]  The variational principle
\[
\delta\int_U\mathcal{L}(j^1s)=0
\]
holds, for variations with compact support.
\item[{\bf (vi)}] The section $s$ satisfies the \textbf{covariant Euler-Lagrange equations} for $\mathcal{L}$.
\item[{\bf (vii)}] The variational principle
\[
\delta\int_U\ell(\bar\sigma(x))=0
\]
holds, using variations of the form
\[
\delta\bar\sigma=\mathbf{d}\eta-[\bar\sigma,\eta],
\]
where $\eta:U\subset X\rightarrow \mathfrak{g}$ has compact support, and $\mathbf{d}$ denotes the derivative on $X$.
\item[{\bf (viii)}] The \textbf{covariant Euler-Poincar\'e equations} hold:
\[
\operatorname{div}\frac{\delta \bar{\ell}}{\delta\bar\sigma}=-\operatorname{Tr}\left(\operatorname{ad}^*_{\bar\sigma}\frac{\delta \bar{\ell}}{\delta\bar\sigma}\right).
\]
where $\operatorname{div}$ denotes the divergence on $X$.
\end{itemize}
\end{theorem}
\textbf{Proof.} The statements ${\bf (i)-(iv)}$ are equivalent by the affine Euler-Poincar\'e reduction theorem. The statements ${\bf (v)-(viii)}$ are equivalent by the covariant Euler-Poincar\'e reduction, where the trivial connection has been chosen. As we have seen before, equations ${\bf(iv)}$ and ${\bf(viii)}$ are equivalent since $\bar{\sigma}^1(x)=\nu(t)(m)$ and $\bar{\sigma}^2(x)=-\gamma(t)(m).\qquad\blacksquare$

\medskip

One can also check that the constrained variational principles ${\bf(iii)}$ and ${\bf(vii)}$ are identical. Indeed, if $\eta(x)=\xi(t)(m)$, we have $\mathbf{d}\eta=\dot\zeta\partial_t+\mathbf{d}\zeta$ and $[\bar\sigma,\eta]=[\nu,\zeta]\partial_t-[\gamma,\zeta]$, thus $\delta\bar\sigma=\mathbf{d}\eta-[\bar\sigma,\eta]$ is equivalent to
\[
\delta\nu=\dot\zeta-[\nu,\zeta]\quad\text{and}\quad\delta\gamma=-\mathbf{d}^\gamma\zeta.
\]
\paragraph{Compatibility condition.} As we have mentioned before, the covariant Euler-Poincar\'e equations are not sufficient for reconstructing the
solution of the original variational problem. One must impose the additional compatibility condition given by the vanishing of the curvature:
\[
\mathbf{d}^{\mathcal{A}^\sigma}\mathcal{A}^\sigma=0,
\]
where $\mathcal{A}^\sigma$ is the connection associated to the reduced jet extension $\sigma$. Recall the in the case of a trivial principal bundle, we have
\[
\overline{\mathcal{A}^\sigma}(x)=-\bar\sigma(x)=-(\nu(t)(m),-\gamma(t)(m))=(-\nu(t)(m),\gamma(t)(m)).
\]
and we get
\[
\mathbf{d}^{\mathcal{A}^\sigma}\mathcal{A}^\sigma\Leftrightarrow \mathbf{d}^{-\bar\sigma}(-\bar\sigma)=0\Leftrightarrow\left\{\begin{array}{l}\mathbf{d}^\gamma\gamma=0\\ \dot\gamma+\mathbf{d}^\gamma\nu=0\end{array}\right.
\]
The compatibility condition of the covariant point of view is equivalent to the affine advection equation $\gamma+\mathbf{d}^\gamma\nu=0$ of the connection $\gamma$ together with the vanishing of the curvature $\mathbf{d}^\gamma\gamma=0$ valid in the particular case $\gamma_0=0$

\section{Affine Euler-Poincar\'e reduction on a principal bundle}
\label{sec:affine_EP_general}

We now generalize the theory developed in Section \ref{spin_systems} to the case of a not necessarily trivial  \textit{right} principal bundle $P_M\rightarrow M$ with structure group $G$. We begin by recalling our conventions and the necessary facts used in the subsequent sections; for details and more information see \cite{GBRa2008a}.  
\medskip

The group of gauge transformations of $P_M$ is the group $\mathcal{G}au(P_M)$ of all diffeomorphisms $\varphi$ of $P_M$ such that $\pi_M\circ\varphi=\pi_M$ and $\varphi\circ\Phi_g=\Phi_g\circ\varphi$. The Lie algebra $\mathfrak{gau}(P_M)$ of $\mathcal{G}au(P_M)$ consists of all $G$-invariant and vertical vector fields $U$ on $P_M$. Thus the $\mathfrak{gau}(P_M)$ can be identified with the Lie algebra $\mathcal{F}_G(P_M,\mathfrak{g})$ of all equivariant maps $\mathcal{U}:P_M\rightarrow \mathfrak{g}$, that is, we have
\[
\mathcal{U}(\Phi_g(p))=\operatorname{Ad}_{g^{-1}}\mathcal{U}(p),\quad\text{for all $g\in G$.}
\]
The Lie algebra isomorphism is given by
\[
\mathcal{U}\in \mathcal{F}_G(P_M,\mathfrak{g})\mapsto U\in\mathfrak{gau}(P_M),\quad U(p):=\left(\mathcal{U}(p)\right)_P(p),
\]
where, for $\xi\in\mathfrak{g}$, $\xi_P$ is the infinitesimal generator defined in \eqref{infinitesimal}. We can also identify $\mathcal{F}_G(P_M,\mathfrak{g})$ with the Lie algebra $\Gamma(\operatorname{Ad}P_M)$ of sections of the adjoint bundle, the identification being given by
\[
\mathcal{U}\in\mathcal{F}_G(P,\mathfrak{g})\mapsto \tilde{\mathcal{U}} \in\Gamma(\operatorname{Ad}P_M),\quad \tilde{\mathcal{U}}(m):=[p,\mathcal{U}(p)]_G.
\]

The dual space to $\mathfrak{gau}(P_M)$ is given by
\[
\mathfrak{gau}(P_M)^*=\left\{\alpha\in\Gamma\left((VP)^*\right)\mid \Phi_g^*\alpha=\alpha\right\},
\]
that is, for all $p\in P$, $\alpha(p)$ is a linear form on the vertical space $V_pP$ and we have
\[
\alpha(\Phi_g(p))(T\Phi_g(u_p))=\alpha(p)(u_p),\quad\text{for all $g\in G$}.
\]
The duality pairing is given by integration, over $M$, of the function $m\mapsto \alpha(p)\!\cdot\!U(p),$ $\pi(p)=m$, that is
\[
\langle\alpha,U\rangle:=\int_M\alpha(p)\!\cdot\!U(p)\mu_M
\]
The dual space $\mathfrak{gau}(P_M)^*$ can be identified with the spaces $\mathcal{F}_G(P_M,\mathfrak{g}^*)$ and $\Gamma(\operatorname{Ad}P_M^*)$ as follows.
\[
\alpha\in \mathfrak{gau}(P_M)^*\mapsto \mu\in \mathcal{F}_G(P_M,\mathfrak{g}^*),\quad \mu(p):=\mathbb{J}(\alpha(p)),
\]
where $\mathbb{J}:T^*P\rightarrow\mathfrak{g}$ is the momentum map of the cotangent lifted action of $G$ on $T^*P$, and
\begin{equation}\label{tilde_mu}
\mu:\mathcal{F}_G(P_M,\mathfrak{g}^*)\mapsto \tilde{\mu}\in \Gamma(\operatorname{Ad}P_M^*),\quad\tilde{\mu}(m):=[p,\mu(p)]_G.
\end{equation}

Denoting by $\Omega^1_G(P_M,\mathfrak{g})$ the space of all $\mathfrak{g}$-valued one-forms on $P_M$ satisfying the condition
\[
\Phi_g^*\omega=\operatorname{Ad}_{g^{-1}}\circ\,\omega
\]
and by $\mathfrak{X}_G(P_M,\mathfrak{g}^*)$ the space of all $\mathfrak{g}^*$-valued vector fields on $P_M$ satisfying
\[
\Phi_g^*X=\operatorname{Ad}^*_g\circ\,X,
\]
we have the $L^2$ duality paring between $\mathfrak{X}_G(P_M,\mathfrak{g}^*)$ and $\Omega^1_G(P_M,\mathfrak{g})$ given by
\[
\langle\omega,X\rangle=\int_M\omega(p)\!\cdot\!X(p)
\]
We will consider the subspace 
$\overline{\Omega^1}(P_M,\mathfrak{g})\subset\Omega^1_G(P_M,\mathfrak{g})$, consisting of all $\mathfrak{g}$-valued one-forms $\omega$ on $P_M$ such that
\[
\Phi_g^*\omega=\operatorname{Ad}_{g^{-1}}\circ\,\omega\quad\text{and}\quad\omega(\xi_P)=0,\quad\text{for all $g\in G$ and $\xi\in\mathfrak{g}$},
\]
and the subspace $\overline{\mathfrak{X}}(P_M,\mathfrak{g}^*)\subset \mathfrak{X}_G(P_M,\mathfrak{g}^*)$, consisting of all $\mathfrak{g}^*$-valued vector fields $X$ on $P_M$ such that
\[
\Phi_g^*X=\operatorname{Ad}^*_g\circ\,X\quad\text{and}\quad X(p)\in L\left([V_pP]^\circ,\mathfrak{g}^*\right),
\]
where $[V_pP]^\circ : = \{ \alpha_p \in T^\ast_pP \mid \alpha_p( \xi_P(p)) = 0\; \text{for all}\; \xi\in \mathfrak{g}\}$.
The $L^2$ duality pairing restricts to these subspaces.

The space $\overline{\Omega^1}(P_M,\mathfrak{g})$ is of special importance since it is the underlying vector space of the affine space $\mathcal{C}onn(P_M)$ of all principal connections on $P_M$. Note that $\overline{\Omega^1}(P_M,\mathfrak{g})$ can be identified with the space $\Gamma(L(TM,\operatorname{Ad}P_M))$ of all sections of the vector bundle $L(TM,\operatorname{Ad}P_M)$, the identification being given by
\[
\omega\in \overline{\Omega^1}(P_M,\mathfrak{g})\mapsto\tilde{\omega}\in \Gamma(L(TM,\operatorname{Ad}P_M)),\quad \tilde{\omega}(m)(v_m):=[p,\omega(p)(u_p)]_G,
\]
where $u_p\in T_pP_M$ is such that $T\pi(u_p)=v_m$. In a similar way, its dual space $\overline{\mathfrak{X}}(P_M,\mathfrak{g}^*)$ is identified with the space $\Gamma(L(T^*M,\operatorname{Ad}P_M^*))$, the identification being given by
\begin{equation}\label{tilde_X}
X\in \overline{\mathfrak{X}}(P_M,\mathfrak{g}^*)\mapsto\tilde{X}\in \Gamma(L(T^*M,\operatorname{Ad}P_M^*)),\quad\tilde{X}(m):=\left[p,X\circ (T_p\pi)^*\right]_G
\end{equation}
\medskip

In the case of a general principal bundle $P_M\rightarrow M$, Theorem \ref{AEP_thm} generalizes as follows.

\begin{itemize}
\item Assume that we have a Lagrangian
\[
L:T\mathcal{G}au(P_M)\times\mathcal{C}onn(P_M)\rightarrow\mathbb{R}
\]
which is \textit{right\/} invariant under the action of $\psi\in\mathcal{G}au(P_M)$ given by
\begin{equation}\label{right_action}
(\varphi,\dot\varphi,\Gamma)\mapsto (\varphi\circ\psi,\dot\varphi\circ\psi,\psi^*\Gamma).
\end{equation}
\item For a fixed $\Gamma_0\in\mathcal{C}onn(P_M)$, we define the Lagrangian $L_{\Gamma_0}:T\mathcal{G}au(P_M)\rightarrow\mathbb{R}$ by $L_{\Gamma_0}(\varphi,\dot\varphi):=L(\varphi,\dot\varphi,\Gamma_0)$. Then $L_{\Gamma_0}$ is right invariant under the lift to $T\mathcal{G}au(P_M)$ of the right action
of $\mathcal{G}au(P_M)_{\Gamma_0}$ on $\mathcal{G}au(P_M)$, where $\mathcal{G}au(P_M)_{\Gamma_0}$ denotes the isotropy group of $\Gamma_0$ with respect to the action \eqref{right_action}.
\item Right invariance of $L$ permits us to define the reduced Lagrangian $l=l(U,\Gamma):\mathfrak{gau}(P_M)\times\mathcal{C}onn(P_M)\rightarrow\mathbb{R}$.
\item For a curve $\varphi_t\in \mathcal{G}au(P_M)$, let $U_t:=\dot\varphi_t\circ\varphi_t^{-1}\in\mathfrak{gau}(P_M)$ and define the curve $\Gamma_t$ as the unique solution of the following affine differential equation with time dependent coefficients
\[
\dot\Gamma+\mathbf{d}^\Gamma \mathcal{U}=0,\quad \Gamma(0)=\Gamma_0
\]
where $\mathbf{d}^\Gamma \mathcal{U}=\mathbf{d}\mathcal{U}+[\Gamma,\mathcal{U}]$ is the covariant derivative. The solution can be written as
\[
\Gamma_t=(\varphi_t)_*\Gamma_0.
\]
\end{itemize}

Below, the functional derivative $\delta l/\delta\mathcal{U}$ is interpreted as an element of $\mathcal{F}_G(P_M,\mathfrak{g}^*)$ and $\delta l/\delta\Gamma$ is interpreted as an element in $\overline{\mathfrak{X}}(P_M,\mathfrak{g}^*)$. The covariant divergences
\[
\operatorname{div}^\Gamma:\overline{\mathfrak{X}}(P_M,\mathfrak{g}^*)\rightarrow\mathcal{F}_G(P_M,\mathfrak{g}^*)\quad\text{and}\quad \operatorname{div}^\Gamma:\Gamma(L(T^*M,\operatorname{Ad}P_M))\rightarrow\Gamma(\operatorname{Ad}P^*)
\]
are defined as minus the $L^2$ adjoint to the the covariant derivatives
\[
\mathbf{d}^\Gamma:\mathcal{F}_G(P_M,\mathfrak{g})\rightarrow\overline{\Omega^1}(P_M,\mathfrak{g})\quad\text{and}\quad\nabla^\Gamma:\Gamma(\operatorname{Ad}P_M)\rightarrow \Gamma(L(TM,\operatorname{Ad}P)),
\]
respectively.

\begin{theorem}\label{AEP_thm_nontrivial} With the preceding notations, the following are equivalent:
\begin{itemize}
\item With $\Gamma_0$ held fixed, Hamilton's variational principle
\[
\delta\int_{t_1}^{t_2}L_{\Gamma_0}(\varphi,\dot\varphi)dt=0,
\]
holds, for variations $\delta\varphi$ of $\varphi$ vanishing at the endpoints.
\item $\varphi$ satisfies the Euler-Lagrange equations for $L_{\Gamma_0}$ on $\mathcal{G}au(P_M)$.
\item The constrained variational principle
\[
\delta\int_{t_1}^{t_2}l(U,\Gamma)dt=0,
\]
holds on $\mathcal{G}au(P_M)\times \mathcal{C}onn(P_M)$, upon using variations of the form
\[
\delta\mathcal{U}=\dot\zeta-[\mathcal{U},\zeta],\quad\delta\Gamma=-\mathbf{d}^\Gamma\zeta,
\]
where $\zeta(t)\in\mathcal{F}_G(P_M,\mathfrak{g})$ vanishes at the endpoints.
\item The \textbf{affine Euler-Poincar\'e equations} hold on $\mathcal{F}_G(P_M,\mathfrak{g})\times\mathcal{C}onn(P_M)$:
\begin{equation}\label{AEP_general}
\frac{\partial }{\partial t}\frac{\delta l}{\delta\mathcal{U}}=-\operatorname{ad}^*_{\mathcal{U}}\frac{\delta l}{\delta\mathcal{U}}+\operatorname{div}^\Gamma\frac{\delta l}{\delta\Gamma}.
\end{equation}
\end{itemize}
\end{theorem}

\section{Covariant to dynamic reduction}
\label{sec:covariant_to_dynamic_general}

This section contains the main results of the paper. Given a field theoretic Lagrangian density defined on the first jet bundle of a not necessarily trivial principal bundle we construct a new family of real valued dynamic Lagrangians on the tangent bundle of the gauge group depending parametrically on a connection. If the Lagrangian density is invariant under the structure group of the principal bundle we will show that the dynamic Lagrangian is gauge group invariant.  We perform covariant and dynamic reduction and show that the resulting reduced equations are equivalent.

\subsection{Splitting of the covariant reduction}\label{Splitting}

Let $P_M\rightarrow M$ be a \textit{right\/} principal bundle with structure group $G$. For simplicity we suppose that $M$ has a volume form $\mu_M$. On the manifold $X=\mathbb{R}\times M$ we consider the volume form $\mu=dt\wedge \mu_M$ and the principal bundle $P_X:=\mathbb{R}\times P_M\rightarrow X$. Note that the first jet bundle $J^1P_X$ is isomorphic to the bundle $VP_M\times_XJ^1P_M$. More precisely, each $\gamma_{(t,p)}\in J^1P_{(t,p)}$ reads
\begin{equation}
\label{gamma_t_p}
\gamma_{(t,p)}:T_t\mathbb{R}\times T_mM\rightarrow T_t\mathbb{R}\times T_pP,\quad \gamma_{(t,p)}((t,v),u_m)=((t,v),a_pv+\gamma_p(u_m)),
\end{equation}
where $a_p\in V_pP$ and $\gamma_p\in J^1P_M$.

Consider a $G$ invariant Lagrangian density $\mathcal{L}:J^1P_X\rightarrow \Lambda^{n+1}X$. Using the previous notation, we can write
\begin{equation}\label{splitting_of_gamma}
\mathcal{L}(\gamma_{(t,p)})=\mathcal{L}(t,a_p,\gamma_p).
\end{equation}
In the same way, the reduced jet bundle $J^1P_X/G$ can be identified with the bundle $\left(VP_M/G\right)\times_X \left(J^1P_M/G \right)$. Recall that the vector bundle $VP_M/G$ is canonically isomorphic to the adjoint bundle $\operatorname{Ad}P_M$, the isomorphism being given by the map $\sigma$ defined by
\[
[p,\xi]_G\in \left(\operatorname{Ad}P_M\right)_x\mapsto \sigma_x\left([p,\xi]_G\right):=[\xi_P(p)]\in \left(VP_M/G\right)_x.
\]
Note that the inverse is
\begin{equation}
\label{sigma_minus_one}
\sigma_x^{-1}[a_p]=[p,\mathcal{A}(a_p)]_G,
\end{equation}
where $\mathcal{A}$ is any principal connection. Note that the right hand side of this formula does not depend on the connection $\mathcal{A}$ since $a _p $ is a vertical vector.
We denote by $\ell$ the reduced Lagrangian density on $J^1P_X/G\cong \left(VP_M/G\right)\times_X \left(J^1P_M/ G \right) \cong \left(\operatorname{Ad}P_M\right)\times_X \left(J^1P_M/G \right)$.

Given a principal connection $\mathcal{A}_M$ on $P_M$, there is a bundle isomorphism $\Psi_{\mathcal{A}_M}:J^1P_M/G\rightarrow L(TM,\operatorname{Ad}P_M)$. Thus we get the connection dependent isomorphism
\[
J^1P_X/G\cong \left(VP_M/G\right)\times_X \left(J^1P_M/G \right)\longrightarrow \left(\operatorname{Ad}P_M\right)\times_XL\left(TM,\operatorname{Ad}P_M\right)
\]
given by
\begin{equation}\label{isomorphism_AM}
[\gamma_{(t,p)}]\cong \left([a_p],[\gamma_p]\right)\longmapsto \left(\sigma_x^{-1}[a_p],[p,\mathcal{A}_M\!\cdot\!\gamma_p]_G\right).
\end{equation}
Note that the connection $\mathcal{A}_M$ on $P_M$ naturally induces a connection $\mathcal{A}$ on $P_X$ given by $\mathcal{A}(t,p)((t,v),u_p):=\mathcal{A}_M(p)(u_p)$. Using the connection $\mathcal{A}$, we have the bundle isomorphism $\Psi_\mathcal{A}: 
J^1P_X/G\rightarrow L(TX,\operatorname{Ad}P)\cong L(TX,\operatorname{Ad}P_M)$ over $X$. This isomorphism is equivalent to \eqref{isomorphism_AM}. Indeed, using \eqref{gamma_t_p} and \eqref{sigma_minus_one}, given $[\gamma_{(t,p)}]\in J^1P_X/G$ and $((t,v),u_m)\in T_xX$, we have
\begin{align*}
\Psi_{\mathcal{A}}([\gamma_{(t,p)}])((t,v),u_m)&=[(t,p),\mathcal{A}(t,p)((t,v),a_pv+\gamma_p(u_m))]_G\\
&=\left[p,\mathcal{A}_M(a_p) v+\mathcal{A}_M(\gamma_p(u_m))\right]_G\\
&=\sigma_x^{-1}[a_p]v+\left[p,\mathcal{A}_M(\gamma_p(u_m))\right]_G.
\end{align*}
As before, we denote by $\ell^\mathcal{A}:=\ell\circ\Psi_{\mathcal{A}}^{-1}$ the reduced Lagrangian induced on
\begin{align*}
L\left(TX,\operatorname{Ad}P_M\right)&\cong L\left(T\mathbb{R},\operatorname{Ad}P_M\right)\times_X L\left(TM,\operatorname{Ad}P_M\right)\\
&\cong \operatorname{Ad}P_M\times_X L\left(TM,\operatorname{Ad}P_M\right)
\end{align*}
by the map $\Psi_\mathcal{A}$.

From now on, $V\subset M$ will denote an open subset of $M$ and $U$ is defined by $U:=I\times V\subset X$, where $I$ is an open interval. Given a local section $s:U\subset X\rightarrow P_X$, we will use the identification
\[
P_X\ni s(x)=(t,s_t(m))\cong s_t(m)\in P_M,\quad x=(t,m)
\]
and we shall regard $s_t$ as a local time-dependent section of $P_M$. Interpreted this way, the first jet extension reads
\[
j^1s(x)=(\dot s_t(m),j^1s_t(m))\in V_{s(x)}P_M\times (J^1P_M)_{s(x)},
\]
where $\dot s$ denotes the partial derivative with respect to the variable $t$, and $j^1s_t$ denotes the first jet extension of the time-dependent section $s_t$ of $P_M$. Note that the relation $\pi_M\circ s_t=id_M$ implies
\[
T\pi_M(\dot s_t(m))=0\quad\text{and}\quad T\pi_M\!\cdot\! j^1s_t(m)=id_{T_mM}.
\]
The reduced first jet extension reads
\begin{align*}
\sigma^\mathcal{A}(x)=&\Psi_{\mathcal{A}}\left(\left[j^1s(x)\right]\right)=\Psi_{\mathcal{A}}\left(\left[\dot s_t(m),j^1s_t(m)\right]\right)\\
=&\sigma_x^{-1}[\dot s_t(m)]+\left[s_t(m),\mathcal{A}_M\!\cdot\!j^1s_t(m)\right]\\
=&:\left(\sigma^1_t(m),\sigma^2_t(m)\right)\in\left(\operatorname{Ad}P_M\right)_m\times_X L\left(T_mM,\left(\operatorname{Ad}P_M\right)_m\right).
\end{align*}
Thus, the reduced jet extension $\sigma^\mathcal{A}(x)$ can be identified with the two time dependent sections $\sigma^1_t$ and $\sigma^2_t$ of the vector bundles
\[
\operatorname{Ad}P_M\rightarrow M\quad\text{and}\quad L\left(TM,\operatorname{Ad}P_M\right)\rightarrow M,
\]
respectively. We can write the reduced Lagrangian as
\[
\ell^\mathcal{A}(\sigma(x))=\ell^\mathcal{A}(t,\sigma^1_t(m),\sigma^2_t(m)),
\]
and we can identify the fiber derivative $\frac{\delta\bar \ell^\mathcal{A}}{\delta\sigma^\mathcal{A}}$ with the time dependent sections
\[
\frac{\delta\bar\ell^\mathcal{A}}{\delta\sigma^1}\quad\text{and}\quad \frac{\delta\bar\ell^\mathcal{A}}{\delta\sigma^2}
\]
of the vector bundles
\[
\operatorname{Ad}P_M^*\rightarrow M\quad\text{and}\quad L\left(T^*M,\operatorname{Ad}P_M^*\right)\rightarrow M,
\]
respectively. Thus, the covariant Euler-Poincar\'e equations \eqref{Cov_EP_A} reads
\begin{equation}\label{splitting_of_cov_EP}
\frac{\partial}{\partial t}\frac{\delta\bar\ell^\mathcal{A}}{\delta\sigma^1}+\operatorname{div}^{\mathcal{A}_M}\frac{\delta\bar\ell^\mathcal{A}}{\delta\sigma^2}=-\operatorname{ad}^*_{\sigma^1}\frac{\delta\bar\ell^\mathcal{A}}{\delta\sigma^1}-\operatorname{Tr}\left(\operatorname{ad}^*_{\sigma^2}\frac{\delta\bar\ell^\mathcal{A}}{\delta\sigma^2}\right).
\end{equation}

\subsection{Definition and gauge invariance of the instantaneous Lagrangian}

Given a $G$-invariant Lagrangian density $\mathcal{L}:J^1P_X\rightarrow\Lambda^{n+1}X$, we write
\[
\mathcal{L}=\overline{\mathcal{L}}\,dt\wedge\mu_M,
\]
and we have
\[
\mathbf{i}_{\partial_t}\mathcal{L}=\overline{\mathcal{L}}\mu_M.
\]
In order to define the instantaneous Lagrangian we shall need the following Lemma.

\begin{lemma} Let $(\varphi,\dot\varphi)\in T\mathcal{G}au(P_M)$, $\mathcal{L}:J^1P_X\rightarrow\Lambda^{n+1}X$ be a $G$-invariant Lagrangian density, and $\Gamma$ be a principal connection on $P_M$. Then for all $t$, the map
\[
p\in P_M\mapsto \mathbf{i}_{\partial_t}\mathcal{L}\left(t,\dot\varphi(p),\operatorname{Hor}_{\varphi(p)}^{\varphi_*\Gamma}\right)\in \Lambda^nM
\]
is well defined and $G$-invariant. Thus, it induces a map
\[
m\in M\mapsto   \mathbf{i}_{\partial_t}\mathcal{L}\left(t,\dot\varphi(p),\operatorname{Hor}_{\varphi(p)}^{\varphi_*\Gamma}\right)\in\Lambda^nM.
\]
\end{lemma}
\textbf{Proof.} To show that the map is well-defined, it suffices to see that
\[
\left(\dot\varphi(p),\operatorname{Hor}_{\varphi(p)}^{\varphi_*\Gamma}\right)\in V_{\varphi(p)}P_M\times J^1_{\varphi(p)}P_M\cong J^1_{ \varphi(p)}P_X.
\]
Thus, it makes sense to evaluate the Lagrangian density $\mathcal{L}$ on it. Note that for all $g\in G$, we have
\[
\mathcal{L}\left(t,\dot\varphi(\Phi_g(p)),\operatorname{Hor}_{\varphi(\Phi_g(p))}^{\varphi_*\Gamma}\right)=\mathcal{L}\left(t,T\Phi_g(\dot\varphi(p)),T\Phi_g\!\cdot\!\operatorname{Hor}_{\varphi(p)}^{\varphi_*\Gamma}\right)=\mathcal{L}\left(t,\dot\varphi(p),\operatorname{Hor}_{\varphi(p)}^{\varphi_*\Gamma}\right).
\]
This proves that for all $t$, the map $m\mapsto \mathcal{L}\left(t,\dot\varphi(p),\operatorname{Hor}_{\varphi(p)}^{\varphi_*\Gamma}\right)$, where $\pi(p)=m$, is well-defined on $M.\qquad\blacksquare$
\medskip

We are now ready to states the main definition of this section.

\begin{definition}\label{definition_L_nontrivial} Let $\mathcal{L}:J^1P_X\rightarrow\Lambda^{n+1}X$ be a $G$-invariant Lagrangian density. The \textbf{instantaneous Lagrangian} is defined by
\[
L^\mathcal{L}=L^\mathcal{L}(t,\varphi,\dot\varphi,\Gamma):I\times T\mathcal{G}au(P_M)\times\mathcal{C}onn(P_M)\rightarrow\mathbb{R},
\]
\begin{equation}\label{definition_L_general_case}
L^\mathcal{L}(\varphi,\dot\varphi,\Gamma):=\int_M\mathbf{i}_{\partial_t}\mathcal{L}\left(t,\dot\varphi(p),\operatorname{Hor}_{\varphi(p)}^{\varphi_*\Gamma}\right).
\end{equation}
\end{definition}

The Lagrangian $L$ has the remarkable property to be invariant under the right action of the gauge transformation group $\mathcal{G}au(P_M)$. This is stated in the following theorem which generalizes Theorem \ref{main_theorem_trivial_bundle} to the case of non-trivial principal bundles.

\begin{theorem} Consider a $G$-invariant Lagrangian density $\mathcal{L}:J^1P\rightarrow \Lambda^{n+1}X$ and its associated Lagrangian defined in \eqref{definition_L_general_case}. Then $L^\mathcal{L}$ is well defined and for all $\psi\in\mathcal{G}au(P_M)$ we have
\[
L^\mathcal{L}(t,\varphi\circ\psi,\dot\varphi\circ\psi,\psi^*\Gamma)=L^\mathcal{L}(t,\varphi,\dot\varphi,\Gamma).
\]
\end{theorem}
\textbf{Proof.} For all $\psi\in\mathcal{G}au(P_M)$ we have
\begin{align*}
L^\mathcal{L}(t,\varphi\circ\psi,\dot\varphi\circ\psi,\psi^*\Gamma)&=\int_M\mathbf{i}_{\partial_t}\mathcal{L}\left(t,\dot\varphi(\psi(p)),\operatorname{Hor}_{\varphi(\psi(p))}^{(\varphi\circ\psi)_*\psi^*\Gamma}\right)\\
&=\int_M\mathbf{i}_{\partial_t}\mathcal{L}\left(t,\dot\varphi(\psi(p)),\operatorname{Hor}_{\varphi(\psi(p))}^{\varphi_*\Gamma}\right).
\end{align*}
Since $\psi$ is a gauge transformation, we can write $\psi(p)=\Phi_{\tau(p)}(p)$ where $\tau:P\rightarrow G$ is such that
\[
\tau(\Phi_g(p))=g^{-1}\tau(p)g,\quad\text{for all $g\in G$}.
\]
We obtain
\[
\varphi(\psi(p))=\varphi(\Phi_{\tau(p)}(p))=\Phi_{\tau(p)}(\varphi(p)),
\]
and we have $\operatorname{Hor}_{\varphi(\psi(p))}^{\varphi_*\Gamma}=T\Phi_{\tau(p)}\!\cdot\!\operatorname{Hor}_{\varphi(p)}^{\varphi_*\Gamma}$. Similarly, we have
\[
\dot\varphi(\psi(p))=T\Phi_{\tau(p)}(\dot\varphi(p)),
\]
thus, by invariance of the Lagrangian density $\mathcal{L}$, we can write
\begin{align*}
L^\mathcal{L}(t,\varphi\circ\psi,\dot\varphi\circ\psi,\psi^*\Gamma)&=\int_M\mathbf{i}_{\partial_t}\mathcal{L}\left(t,T\Phi_{\tau(p)}(\dot\varphi(p)),T\Phi_{\tau(p)}\left(\operatorname{Hor}_{\varphi(p)}^{\varphi_*\Gamma}\right)\right)\\
&=\int_M\mathbf{i}_{\partial_t}\mathcal{L}\left(t,\dot\varphi(p),\operatorname{Hor}_{\varphi(p)}^{\varphi_*\Gamma}\right)\\
&=L^\mathcal{L}(t,\varphi,\dot\varphi,\Gamma).\qquad\blacksquare
\end{align*}

Recall from Theorem \ref{AEP_thm_nontrivial} that given a possibly time dependent gauge invariant Lagrangian
\[
L:T\mathcal{G}au(P_M)\times\mathcal{C}onn(P_M)\rightarrow\mathbb{R}
\]
and a fixed connection $\Gamma_0\in\mathcal{C}onn(P_M)$, a curve $\varphi\in\mathcal{G}au(P_M)$ is a solution of the Euler-Lagrange equation for $L_{\Gamma_0}$ if and only if $U:=\dot\varphi\circ\varphi^{-1}\in\mathfrak{gau}(P_M)$ and $\Gamma:=\varphi^*\Gamma_0$ are solutions of the affine Euler-Poincar\'e equations
\[
\frac{\partial}{\partial t}\frac{\delta l}{\delta \mathcal{U}}=-\operatorname{ad}^*_\mathcal{U}\frac{\delta l}{\delta \mathcal{U}}+\operatorname{div}^\Gamma \frac{\delta l}{\delta \Gamma},
\]
where $l:I\times\mathfrak{gau}(P_M)\times\mathcal{C}onn(P_M)\rightarrow\mathbb{R}$ is the reduced Lagrangian induced by $L$. 

By $G $-invariance of $\mathcal{L}$ and $\overline{\mathcal{L}}$ and $\mathbf{i}_{\partial_t}\mathcal{L}=\overline{\mathcal{L}}\mu_M$, we deduce that $\mathbf{i}_{\partial_t} \ell = \bar{ \ell} \mu_M$. Thus, in
the case of $L^\mathcal{L}$, the reduced Lagrangian is given by
\begin{align*}\label{reduced_Lagr}
l^\mathcal{L}(t,U,\Gamma)&=\int_M\mathbf{i}_{\partial_t}\mathcal{L}\left(t,U(p),\operatorname{Hor}^\Gamma_p\right)\\
&=\int_M\bar\ell\Big(t,[U(p)],\left[\operatorname{Hor}^\Gamma_p\right]\Big)\mu_M\\
&=\int_M\bar\ell^\mathcal{A}\Big(t,\tilde{\mathcal{U}}(m),\left[p,\mathcal{A}_M\!\cdot\!\operatorname{Hor}^\Gamma_p\right]_G\Big)\mu_M,
\end{align*}
where $\bar\ell$ is the reduced Lagrangian density on $J^1P_X/G\cong \left(VP_M/G\right)\times_X \left(J^1P_M/G\right)$ and $\bar\ell^\mathcal{A}:=\ell\circ\Psi_{\mathcal{A}}^{-1}$ is the reduced Lagrangian density on $\operatorname{Ad}P_M\times_X L\left(TM,\operatorname{Ad}P_M\right)$. Recall that $\ell^\mathcal{A}$ is defined with the help of a fixed connection $\mathcal{A}_M$ on $P_M$. We will always choose
\[
\mathcal{A}_M=\Gamma_0,
\]
where $\Gamma_0$ is the initial value of the connection in the affine Euler-Poincar\'e picture. We thus have
\[
l^\mathcal{L}(t,U,\Gamma)=\int_M\bar\ell^{\,\Gamma_0}\Big(t,\tilde{\mathcal{U}}(m),\left[p,\Gamma_0\!\cdot\!\operatorname{Hor}^\Gamma_p\right]_G\Big)\mu_M.
\]

\subsection{Affine Euler-Poincar\'e formulation of covariant reduction}

The situation described in the preceding two subsections can be summarized in the following commutative diagram:
\[
\xymatrix{
\mathcal{L}:J^1P_X\rightarrow\Lambda^{n+1}X\quad \ar[r] \ar[d]_{\text{reduction by $G$}\;}& L^{\mathcal{L}}:I\times T\mathcal{G}au(P_M)\times\mathcal{C}onn(P_M)\rightarrow\mathbb{R}\ar [d]^{\;\text{reduction by $\mathcal{G}au(P_M)$}}\\ \ell:J^1P_X/G\rightarrow \Lambda^{n+1}X\quad \ar[r] & l^{\mathcal{L}}: I\times \mathfrak{gau}(P_M)\times\mathcal{C}onn(P_M)\rightarrow\mathbb{R}.
}
\]
Given a $G$-invariant Lagrangian density $\mathcal{L}:J^1P_X\rightarrow\Lambda^{n+1}X$, we associate the gauge invariant time dependent Lagrangian
\[
L^\mathcal{L}(t,\varphi,\dot\varphi,\Gamma_0):=\int_M\mathbf{i}_{\partial_t}\mathcal{L}\left(t,\dot\varphi(p),\operatorname{Hor}_{\varphi(p)}^{\varphi_*\Gamma_0}\right).
\]
To the reduced Lagrangian density $\ell:J^1P_X/G\rightarrow\Lambda^{n+1}X$ we associate the time dependent Lagrangian
\begin{equation}\label{Lagrangians}
l^\mathcal{L}(t,U,\Gamma):=\int_M\mathbf{i}_{\partial_t}\ell\left(t,[U(p)],\left[\operatorname{Hor}^\Gamma_p\right]\right)=\int_M\mathbf{i}_{\partial_t}\ell^{\Gamma_0}\left(t,\tilde{\mathcal{U}}(m),\left[p,\Gamma_0\!\cdot\!\operatorname{Hor}^\Gamma_p\right]_G\right).
\end{equation}
Then $l^\mathcal{L}$ is precisely the Lagrangian obtained by reduction of $L^\mathcal{L}$. In this sense we say that the diagram commutes. We now show the link between the functional derivatives of $\ell^{\Gamma_0}$ and $l^\mathcal{L}$.
Recall from Section \ref{sec:affine_EP_general} that $\frac{\delta l^\mathcal{L}}{\delta\mathcal{U}}\in\mathcal{F}_G(P_M,\mathfrak{g}^*)$ and $\frac{\delta l^\mathcal{L}}{\delta\Gamma}\in\overline{\mathfrak{X}}(P_M,\mathfrak{g}^*)$. The functional derivatives of $\ell^{\Gamma_0}$ are simply fiber derivatives.

\begin{lemma}\label{link_functional} Consider the Lagrangian $l^\mathcal{L}$ and $\ell^{\Gamma_0}$ as defined in \eqref{Lagrangians}. Then we have the relations
\[
\widetilde{\frac{\delta l^\mathcal{L}}{\delta\mathcal{U}}}=\frac{\delta \ell^{\Gamma_0}}{\delta\sigma^1}\quad\text{and}\quad \widetilde{\frac{\delta l^\mathcal{L}}{\delta\Gamma}}=-\frac{\delta \ell^{\Gamma_0}}{\delta\sigma^2},
\]
where we use the isomorphisms $\widetilde{\quad}$ defined in \eqref{tilde_mu} and \eqref{tilde_X}, respectively.
\end{lemma}
\textbf{Proof.} We have
\begin{align*}
\int_M\widetilde{\frac{\delta l^\mathcal{L}}{\delta\mathcal{U}}}(m)\!\cdot\!\tilde{\mathcal{V}}(m)\mu_M&=\int_M\frac{\delta l^\mathcal{L}}{\delta\mathcal{U}}(p)\!\cdot\!\mathcal{V}(p)\mu_M=\left.\frac{d}{d\varepsilon}\right|_{\varepsilon=0}l^\mathcal{L}(t,U+\varepsilon V,\Gamma)\\
&=\left.\frac{d}{d\varepsilon}\right|_{\varepsilon=0}\int_M\ell^{\Gamma_0}\left(t,\tilde{\mathcal{U}}(m)+\varepsilon\tilde{\mathcal{V}}(m),\left[p,\Gamma_0\!\cdot\!\operatorname{Hor}^\Gamma_p\right]_G\right)\mu_M\\
&=\int_M\frac{\delta\ell^{\Gamma_0}}{\delta\sigma^1}(m)\!\cdot\!\tilde{\mathcal{V}}(m)\mu_M,
\end{align*}
this proves the first equality. We will use the formula
\[
\left.\frac{d}{d\varepsilon}\right|_{\varepsilon=0}\operatorname{Hor}^{\Gamma+\varepsilon \omega}_p(v_m)=-\left(\omega\!\cdot\!\operatorname{Hor}^\Gamma_p(v_m)\right)_P(p)\in V_pP_M,
\]
valid for all $\omega\in\overline{\Omega^1}(P_M,\mathfrak{g})$, the tangent space to $\mathcal{C}onn(P_M)$. We have 
\begin{align*}
\int_M\widetilde{\frac{\delta l^\mathcal{L}}{\delta\Gamma}}(m)\!\cdot\!\tilde{\omega}(m)\mu_M&=\int_M\frac{\delta l^\mathcal{L}}{\delta\Gamma}(p)\!\cdot\!\omega(p)\mu_M=\left.\frac{d}{d\varepsilon}\right|_{\varepsilon=0}l^\mathcal{L}(t,U,\Gamma+\varepsilon \omega)\\
&=\left.\frac{d}{d\varepsilon}\right|_{\varepsilon=0}\int_M\ell^{\Gamma_0}\left(t,\tilde{\mathcal{U}}(m),\left[p,\Gamma_0\!\cdot\!\operatorname{Hor}^{\Gamma+\varepsilon\omega}_p\right]_G\right)\mu_M\\
&=\int_M\frac{\delta\ell^{\Gamma_0}}{\delta\sigma^2}(m)\!\cdot\!\left.\frac{d}{d\varepsilon}\right|_{\varepsilon=0}\left[p,\Gamma_0\!\cdot\!\operatorname{Hor}^{\Gamma+\varepsilon\omega}_p\right]_G\mu_M\\
&=-\int_M\frac{\delta\ell^{\Gamma_0}}{\delta\sigma^2}(m)\!\cdot\!\left[p,\Gamma_0\left(\left(\omega\!\cdot\!\operatorname{Hor}^\Gamma_p(v_m)\right)_P(p)\right)\right]_G\mu_M\\
&=-\int_M\frac{\delta\ell^{\Gamma_0}}{\delta\sigma^2}(m)\!\cdot\!\left[p,\omega\!\cdot\!\operatorname{Hor}^\Gamma_p(v_m)\right]_G\mu_M\\
&=-\int_M\frac{\delta\ell^{\Gamma_0}}{\delta\sigma^2}(m)\!\cdot\!\tilde\omega(m)\mu_M.
\end{align*}
This proves the second equality.$\quad\blacksquare$

\medskip

Note that the correspondence of the various Lagrangians holds globally, that is, on the whole base manifolds $X$ and $M$. The situation is different for the dynamics, since a principal bundle admits only local sections unless the bundle is trivial.

Therefore, from now on, we need to fix an open subset $V\subset M$ and an equivariant map $\xi_0:P_V\rightarrow G$ such that
\[
\xi_0(\Phi_g(p))=\xi_0(p)g.
\]
This is equivalent to the  choice of a preferred local section $s_0:V\rightarrow P_V$, given by
\[
s_0(m)=\Phi_{\xi_0(p)^{-1}}(p),
\]
where $p$ is such that $\pi(p)=m$. To the map $\xi_0$ we associate the principal connection $\Gamma_0$ on $P_V\rightarrow V$ by
\[
\Gamma_0=TL_{\xi_0^{-1}}T\xi_0.
\]
We choose an open subset $I\subset\mathbb{R}$ and define $U:=I\times V\subset X$. In order to prove the main result of this section, we will need the following technical lemma.

\begin{lemma}\label{technical_lemma} Let $\xi_0:P_V\rightarrow G$ be an equivariant map and $\Gamma_0$ be the associated principal connection as defined above. Let $s:U\subset X\rightarrow P_U\subset P_X$ be a local section and $\varphi_t$ a curve in $\mathcal{G}au(P_V)$, and suppose that they verify the relation
\[
s_t(m)=\Phi_{\xi_0(p)^{-1}}(\varphi_t(p)).
\]
Then we have
\[
\dot s_t(m)=T\Phi_{\xi_0(p)^{-1}}\left(\dot\varphi_t(p)\right)\quad\text{and}\quad j^1s_t(m)=T\Phi_{\xi_0(p)^{-1}}\!\cdot\!\operatorname{Hor}^{(\varphi_t)_*\Gamma_0}_{\varphi_t(p)}.
\]
The second equality can be rewritten as
\[
j^1s_t(m)=T\Phi_{\tau_t(p)\xi_0(p)^{-1}}\left(\operatorname{Hor}_p^{\Gamma_0}+\left(TR_{\tau_t(p)^{-1}}T_p\tau_t\!\cdot\!\operatorname{Hor}_p^{\Gamma_0}\right)_P(p)\right),
\]
where the curve $\tau_t\in\mathcal{F}_G(P_V,G)$ is defined by
\[
\varphi_t(p)=\Phi_{\tau_t(p)}(p).
\]
Define the reduced objects $\sigma^{\Gamma_0}=(\sigma_t^1,\sigma_t^2)$, $U$, and $\Gamma$ by
\[
\sigma_t^1(m)=\sigma_x^{-1}[\dot s_t(m)],\quad \sigma_t^2(m)=\left[s_t(m),\Gamma_0\!\cdot\!j^1s_t(m)\right]_G,\quad U_t=\dot\varphi_t\circ\varphi_t^{-1},\quad\text{and}\quad\Gamma_t=(\varphi_t)_*\Gamma_0.
\]
Then, we have the relations
\[
\sigma_t^1(m)=\sigma_x^{-1}\left[U_t(p)\right]=\tilde{\mathcal{U}}_t(m)\quad\text{and}\quad\sigma_t^2(m)=\left[p,\Gamma_0\!\cdot\!\operatorname{Hor}_p^{\Gamma_t}\right]_G.
\]
The second equality can be rewritten as
\[
\sigma_t^2(m)=\left[p,TR_{\tau_t(p)^{-1}}T\tau_t\!\cdot\!\operatorname{Hor}_p^{\Gamma_0}\right]_G.
\]
\end{lemma}
\textbf{Proof.} The first equality is clearly true. For the second, by differentiating the relation $s_t\circ\pi=\Phi_{\xi_0^{-1}}\circ\varphi_t$ we get
\begin{align*}
Ts_t(T\pi(v_p))&=T\Phi_{\xi_0(p)^{-1}}\left(T\varphi_t(v_p)-\left(TL_{\xi_0(p)^{-1}}T\xi_0(v_p)\right)_P(p)\right)\\
&=T\Phi_{\xi_0(p)^{-1}}\left(T\varphi_t(v_p)-\left(\Gamma_0(v_p)\right)_P(p)\right).
\end{align*}
By choosing $v_p=\operatorname{Hor}^{\Gamma_0}_p(v_m)$ where $v_m\in T_mM$, we obtain
\[
Ts_t(v_m)=T\Phi_{\xi_0(p)^{-1}}\left(T\varphi_t\left(\operatorname{Hor}^{\Gamma_0}_p(v_m)\right)\right)=T\Phi_{\xi_0(p)^{-1}}\left(\operatorname{Hor}^{(\varphi_t)_*\Gamma_0}_{\varphi_t(p)}(v_m)\right).
\]
To obtain the third equality we note that
\[
\operatorname{Hor}^{(\varphi_t)_*\Gamma_0}_{\varphi_t(p)}=T\varphi_t\!\cdot\!\operatorname{Hor}^{\Gamma_0}_p=T\Phi_{\tau_t(p)}\!\cdot\!\left(\operatorname{Hor}^{\Gamma_0}_p+\left(TR_{\tau_t(p)^{-1}}T\tau_t\!\cdot\!\operatorname{Hor}^{\Gamma_0}_p\right)_P(p)\right).
\]
Concerning the reduced objects, we have
\begin{align*}
\sigma_t^1(m)&=\sigma_x^{-1}\left[\dot s_t(m)\right]=\sigma_x^{-1}\left[T\Phi_{\xi_0(p)^{-1}}(\dot\varphi_t(p))\right]=\sigma_x^{-1}\left[U_t(\varphi_t(p))\right]\\
&=\sigma_x^{-1}\left[U_t(\Phi_{\tau_t(p)}(p))\right]=\sigma_x^{-1}\left[T\Phi_{\tau_t(p)}U_t(p)\right]=\sigma_x^{-1}\left[U_t(p)\right]=\widetilde{\mathcal{U}}_t(m)
\end{align*}
and
\begin{align*}
\sigma_t^2(m)&=\left[s_t(m),\Gamma_0\!\cdot\!j^1s_t(m)\right]_G=\left[\Phi_{\xi_0(p)^{-1}}(\varphi_t(p)),\Gamma_0\!\cdot\!T\Phi_{\xi_0(p)^{-1}}\!\cdot\!\operatorname{Hor}^\Gamma_{\varphi_t(p)}\right]_G\\
&=\left[\Phi_{\tau_t(p)\xi_0(p)^{-1}}(p),\Gamma_0\!\cdot\!T\Phi_{\tau_t(p)\xi_0(p)^{-1}}\!\cdot\!\operatorname{Hor}^\Gamma_{p}\right]_G\\
&=\left[p,\Gamma_0\!\cdot\!\operatorname{Hor}^\Gamma_p\right]_G.
\end{align*}
For the last equality, we note that
\begin{align*}
\sigma_t^2(m)&=\left[s_t(m),\Gamma_0\!\cdot\!j^1s_t(m)\right]_G\\
&=\left[\Phi_{\tau_t(p)\xi_0(p)^{-1}}(p),\Gamma_0\!\cdot\!T\Phi_{\tau_t(p)\xi_0(p)^{-1}}\left(\operatorname{Hor}^{\Gamma_0}_p+\left(TR_{\tau_t(p)^{-1}}T\tau_t\!\cdot\!\operatorname{Hor}_p^{\Gamma_0}\right)_P(p)\right)\right]_G\\
&=\left[p,TR_{\tau_t(p)^{-1}}T\tau_t\!\cdot\!\operatorname{Hor}_p^{\Gamma_0}\right]_G. \qquad \qquad\blacksquare
\end{align*}
\medskip

We are now ready to state the main result of the paper.

\begin{theorem}\label{main_result} Consider a right principal $G$-bundle $P_M\rightarrow M$ and define the principal $G$-bundle $P_X:=\mathbb{R}\times P_M\rightarrow X:=\mathbb{R}\times M$ over spacetime.

Let $\mathcal{L}:J^1P_X\rightarrow\Lambda^{n+1}X$ be a $G$-invariant Lagrangian density. Define the associated reduced Lagrangian $\ell$ and the associated Lagrangians $L^\mathcal{L}$ and $l^\mathcal{L}$.

Let $s:U=I\times V\rightarrow P_U=\mathbb{R}\times P_V$ be a local section, choose an equivariant map $\xi_0\in\mathcal{F}_G(V,G)$, define the curve
\[
\varphi_t:=\Phi_{\xi_0}\circ s_t\circ\pi\in\mathcal{G}au(P_V),
\]
and consider the reduced quantities
\[
\sigma^1_t=\sigma_x^{-1}[\dot s_t],\quad \sigma^2_t=[s_t,\Gamma_0\!\cdot\!j^1s_t]_G
\]
\[
U_t=\dot\varphi_t\circ\varphi_t^{-1},\quad\Gamma=(\varphi_t)_*\Gamma_0,
\]
where $\Gamma_0:=TL_{\xi_0^{-1}}T\xi_0$.

Then the following are equivalent.
\begin{itemize}
\item[{\bf (i)}] Hamilton's variational principle
\[
\delta\int_{t_1}^{t_2}L^\mathcal{L}_{\Gamma_0}(t,\varphi_t,\dot\varphi_t)dt=0,
\]
for all variations $\delta\varphi$ of $\varphi$ vanishing at the endpoints.
\item[{\bf (ii)}] The curve $\varphi_t$ satisfies the Euler-Lagrange equations for $L_{\Gamma_0}$ on the gauge group $\mathcal{G}au(P_V)$.
\item[{\bf (iii)}] The constrained affine Euler-Poincar\'e variational principle
\[
\delta\int_{t_1}^{t_2}l^\mathcal{L}(t,U_t,\Gamma_t)dt=0,
\]
holds on $\mathfrak{gau}(P_V)\times\mathcal{C}onn(P_V)$, upon using variations of the form
\[
\delta\mathcal{U}=\dot\zeta-[\mathcal{U},\zeta],\quad\delta\Gamma=-\mathbf{d}^\Gamma\zeta,
\]
where $\zeta_t\in \mathcal{F}_G(P_V,\mathfrak{g})$ vanishes at the endpoints.
\item[{\bf (iv)}] The \textbf{affine Euler-Poincar\'e equations} hold on $\mathfrak{gau}(P_V)\times\mathcal{C}onn(P_V)$:
\[
\frac{\partial}{\partial_t}\frac{\delta l^\mathcal{L}}{\delta\mathcal{U}}=-\operatorname{ad}^*_{\mathcal{U}}\frac{\delta l^\mathcal{L}}{\delta\mathcal{U}}+\operatorname{div}^\Gamma\frac{\delta l^\mathcal{L}}{\delta\Gamma}.
\]
\item[{\bf (v)}] The variational principle
\[
\delta\int_U\mathcal{L}(j^1s)=0
\]
holds, for variations with compact support.
\item[{\bf (vi)}] The local section $s$ satisfies the \textbf{covariant Euler-Lagrange equations} for $\mathcal{L}$.
\item[{\bf (vii)}] The variational principle
\[
\delta\int_U\ell(\sigma)=0
\]
holds, using variations of the form
\[
\delta\sigma=\nabla^{\Gamma_0}\eta-\left[\sigma^{\Gamma_0},\eta\right],
\]
where $\eta:U\subset X\rightarrow\operatorname{Ad}P_U$ is a section with compact support, and
\[
\nabla^{\Gamma_0}:\Gamma(\operatorname{Ad}P_U)\rightarrow\Gamma\left(L\left(TU,\operatorname{Ad}P_U\right)\right)
\]
is the covariant derivative associate to the connection $\Gamma_0$, viewed as a connection on $P_U$.
\item[{\bf (viii)}] The \textbf{covariant Euler-Poincar\'e equations} hold:
\[
\operatorname{div}^{\Gamma_0}\frac{\delta\bar\ell}{\delta\sigma}=-\operatorname{Tr}\left(\operatorname{ad}^*_{\sigma^{\Gamma_0}}\frac{\delta\bar\ell}{\delta\sigma}\right),
\]
where $\operatorname{div}^{\Gamma_0}$ is the covariant divergence associated to the connection $\Gamma_0$ viewed as a connection on $P_U$.
\end{itemize}
\end{theorem}
\textbf{Proof.} The statements $({\bf i}) - ({\bf iv})$ are equivalent by the affine Euler-Poincar\'e reduction theorem. The statements $({\bf v}) - ({\bf viii})$ are equivalent by covariant Euler-Poincar\'e reduction. We now prove that $({\bf iv})$ and $({\bf viii})$ are equivalent. In \S\ref{Splitting}, we have shown that the covariant Euler-Poincar\'e equation $({\bf viii})$ can be rewritten as
\[
\frac{\partial}{\partial t}\frac{\delta\bar\ell^{\,\Gamma_0}}{\delta\sigma^1}+\operatorname{div}^{\Gamma_0}\frac{\delta\bar\ell^{\,\Gamma_0}}{\delta\sigma^2}=-\operatorname{ad}^*_{\sigma^1}\frac{\delta\bar\ell^{\,\Gamma_0}}{\delta\sigma^1}-\operatorname{Tr}\left(\operatorname{ad}^*_{\sigma^2}\frac{\delta\bar\ell^{\,\Gamma_0}}{\delta\sigma^2}\right).
\]
Using Lemma \ref{link_functional} and \ref{technical_lemma}, this equation can be rewritten as
\begin{equation}\label{intermediate_equation}
\frac{\partial}{\partial t}\frac{\delta l^\mathcal{L}}{\delta\mathcal{U}}-\operatorname{div}^{\Gamma_0}\frac{\delta l^\mathcal{L}}{\delta\Gamma}=-\operatorname{ad}^*_{\mathcal{U}}\frac{\delta l^\mathcal{L}}{\delta\mathcal{U}}+\operatorname{Tr}\left(\operatorname{ad}^*_{\sigma^2}\frac{\delta l^\mathcal{L}}{\delta\Gamma}\right).
\end{equation}
Let $\rho$ be a section of the adjoint bundle. It can be written as $\rho(m)=[p,f(p)]_G$ where $f\in \mathcal{F}_G(P_V,\mathfrak{g})$. We have
\begin{align*} \nabla^{\Gamma_0}_{v_m}s&=\left[p,\mathbf{d}f(p)(u_p)+[\Gamma_0(u_p),f(p)]\right]_G,\\ \nabla^{\Gamma}_{v_m}s&=\left[p,\mathbf{d}f(p)(u_p)+[\Gamma(u_p),f(p)]\right]_G,\\  [\sigma_2,s](v_m)&=\left[p,\left[\Gamma_0\left(\operatorname{Hor}^\Gamma_p(v_m)\right),f(p)\right]\right]_G,
\end{align*}
where $u_p$ is such that $T\pi(u_p)=v_m$. By choosing $u_p=\operatorname{Hor}^\Gamma_p(v_m)$ we obtain the equality
\[
\nabla^{\Gamma_0}s=\nabla^{\Gamma}s+[\sigma^2,s],
\]
thus, by duality, we get
\[
-\operatorname{div}^{\Gamma_0}\frac{\delta l^\mathcal{L}}{\delta\Gamma}=-\operatorname{div}^{\Gamma}\frac{\delta l^\mathcal{L}}{\delta\Gamma}+\operatorname{Tr}\left(\operatorname{ad}^*_{\sigma^2}\frac{\delta l^\mathcal{L}}{\delta\Gamma}\right).
\]
Using this equality, equation \eqref{intermediate_equation} can be rewritten as
\[
\frac{\partial}{\partial t}\frac{\delta l^\mathcal{L}}{\delta\mathcal{U}}=-\operatorname{ad}^*_{\mathcal{U}}\frac{\delta l^\mathcal{L}}{\delta\mathcal{U}}+\operatorname{div}^{\Gamma}\frac{\delta l^\mathcal{L}}{\delta\Gamma},
\]
which is exactly the affine Euler-Poincar\'e equation $({\bf iv}).\qquad\blacksquare$

\section{Examples: sigma models and spin glasses}

In this section we apply the theory developed in this paper to \textit{sigma models} over non-trivial principal bundles, that is, the case of a Lagrangian density of the form
\[
\mathcal{L}(\gamma_p)= \frac{1}{2}\|\gamma_p\|^2\mu_X, \quad \gamma_p \in J ^1_pP, 
\]
where the norm is associated to a right invariant bundle metric $K$ on $J^1P$ over $P$. In doing this we have used the affine bundle isomorphism $\mathcal{F}_{\mathcal{A}}: J ^1P \rightarrow L(TX, VP)$ covering $P $ given in \eqref{F_A_def} and the vector bundle structure of $L(TX, VP) \rightarrow P$. This metric can be constructed using a pseudo-Riemannian metric $g$ on $X$, an adjoint invariant inner product $\gamma$ on $\mathfrak{g}$, and a connection $\mathcal{A}$ on $P$ as follows
\[
K(\gamma^1_p,\gamma^2_p):= \operatorname{Tr}_g\left(\gamma\left(\mathcal{A}\!\cdot\!\gamma^1_p,\mathcal{A}\!\cdot\!\gamma^2_p\right)\right),
\]
where $\operatorname{Tr}_g$ is the trace of a bilinear form with respect to the metric $g$. Assuming that we have the trivial slicing $X=\mathbb{R}\times M$ and $P=\mathbb{R}\times P_M$, we can choose the pseudo-Riemannian metric $g=dt^2-g_M$, where $g_M$ is a Riemannian metric on $M$, and the connection $\mathcal{A}((t,v),u_p):=\mathcal{A}_M(u_p)$ as before. In this case, the Lagrangian density reads
\[
\mathcal{L}(a_p,\gamma_p)=\frac{1}{2}\left(\|a_p\|^2-\|\gamma_p\|^2\right)dt\wedge\mu_M,
\]
where the first norm is associated to the metric induced on $VP_M$ by $\gamma$ and the second norm is associated to $g_M, \gamma$, and $\mathcal{A}_M$. The reduced Lagrangian density $\ell^\mathcal{A}: \operatorname{Ad}P_M\times_X L\left(TM,\operatorname{Ad}P_M\right) \rightarrow \Lambda^{n+1}X$ is
\begin{equation}\label{reduced_density_sigma_models}
\ell^\mathcal{A}\left(\sigma^1_x,\sigma^2_x\right)=\frac{1}{2}\left(\|\sigma^1_x\|^2-\|\sigma^2_x\|^2\right)dt\wedge \mu_M.
\end{equation}
Note that, a priori, two principal connections are needed for this example. The first one appears in the expression of the norm in $\mathcal{L}$ and is part of the physical problem. The second is needed to identify the reduced jet bundle with the bundle $\operatorname{Ad}P_M\times_X L\left(TM,\operatorname{Ad}P_M\right)$ through the map $\Psi_\mathcal{A}$. We have of course chosen the same connection in both places. In this case, even if the Lagrangian $\ell^\mathcal{A}$ is constructed from the connection dependent isomorphism $\Psi_\mathcal{A}$, its expression does not depend on the connection, since the norms only involve the metric $g_M$ and the inner product $\gamma$. The covariant Euler-Poincar\'e equation \eqref{splitting_of_cov_EP} associated to the reduced Lagrangian density \eqref{reduced_density_sigma_models} for sigma models becomes
\begin{equation}\label{cov_sigma_models}
\frac{d}{dt}\sigma^1-\operatorname{div}^{\mathcal{A}_M}\sigma^2=\operatorname{Tr}\left(\operatorname{ad}^*_{\sigma^2}\sigma^2\right).
\end{equation}
Recall that $\sigma^1_t$ and $\sigma_t^2$ are time-dependent local sections of the vector bundles $\operatorname{Ad}P_M$ and $L\left(TM,\operatorname{Ad}P_M\right)$ respectively. In writing this equation we have made certain obvious identifications using the metrics on the bundles involved in the computation of the functional derivatives. Thus $\delta \ell^ \mathcal{A}/ \delta \sigma^1 \in \operatorname{Ad}P_M ^*$ but we interpret it as an element of $\operatorname{Ad}P_M $ by using the bundle metric induced by $\gamma $ on $\operatorname{Ad}P_M $. A direct computation then shows that $\delta \ell^ \mathcal{A}/ \delta \sigma^1$ is identified with $\sigma^1 $. Similarly, $\delta \ell^ \mathcal{A}/ \delta \sigma^2 \in L\left(T ^\ast M,\operatorname{Ad}P_M^*\right)$ is identified with an element of $L\left(T M,\operatorname{Ad}P_M\right)$ by using the bundle metric on $L\left(T M,\operatorname{Ad}P_M\right)$ induced by $\gamma $ and $g_M $. A direct computation then shows that $\delta \ell^ \mathcal{A}/ \delta \sigma^2$ is identified with $-\sigma^2$.

According to Definition \ref{definition_L_nontrivial}, the instantaneous Lagrangian associated to $\mathcal{L}$ is
\[
L^\mathcal{L}(t,\varphi,\dot\varphi,\Gamma)=\frac{1}{2}\int_M\left(\|\dot\varphi(p)\|^2-\|\operatorname{Hor}^{\varphi_*\Gamma}_{\varphi(p)}\|^2\right)\mu_M.
\]
From the general theory we know that this Lagrangian is gauge invariant and that the reduced Lagrangian is given by
\[
l^\mathcal{L}(U,\Gamma)=\frac{1}{2}\int_M\left(\|\tilde{\mathcal{U}}(m)\|^2-\|[p,\mathcal{A}_M\!\cdot\!\operatorname{Hor}^\Gamma_p]_G\|^2\right)\mu_M.
\]
The affine Euler-Poincar\'e equations \eqref{AEP_general} become
\begin{equation}\label{dyn_sigma_models}
\frac{d}{dt}\mathcal{U}_t=-\operatorname{div}^{\Gamma_t}\left(\mathcal{A}_M\!\cdot\!\operatorname{hor}^{\Gamma_t}_p\right),
\end{equation}
since we have
\[
\frac{\delta l^\mathcal{L}}{\delta\Gamma}=-\mathcal{A}_M\!\cdot\!\operatorname{hor}^\Gamma_p\in\overline{\Omega^1}(P_V,\mathfrak{g}),
\]
where $\operatorname{hor}^\Gamma$ denotes the horizontal part with respect to the connection $\Gamma$. Recall that here $\mathcal{U}_t$ and $\Gamma_t$ are time dependent curves in $\mathcal{F}_G(P_V,\mathfrak{g})$ and $\mathcal{C}onn(P_V)$, respectively. Here we made the same identifications using various bundle metrics in the calculations of the functional derivatives.

In order to apply the general theory developed previously, we suppose that the connection $\mathcal{A}_M$ appearing in the norm of the sigma model is of the form $\mathcal{A}_M=TL_{\xi_0^{-1}}T\xi_0$, where $\xi_0$ is an equivariant map in $\mathcal{F}_G(P_V,G)$ and we apply the affine Euler-Poincar\'e reduction to the Lagrangian $L_{\Gamma_0}$, where $\Gamma_0=\mathcal{A}_M$. By Theorem \ref{main_result}, the covariant and dynamic reduced equations \eqref{cov_sigma_models} and \eqref{dyn_sigma_models} for sigma models are equivalent. 

Remarkably, in the case of a trivial bundle, we recover the theory of \textit{spin glasses} as described in \cite{Dz1980}. In particular, we obtain the motion equations via covariant Euler-Poincar\'e reduction and the dynamic approach agree with that described in \cite{GBRa2008b} in the case of general spin systems. We indicate briefly how the Lagrangians simplify in the case of a trivial bundle.

Consider the trivial principal bundle $P=X\times G\rightarrow X$ and the Lagrangian density $\mathcal{L}(j^1s):=\|T\bar s\|^2$, where the norm is associated to the right invariant bundle metric on $J^1P$ constructed from a spacetime metric $g=dt^2-g_M$ on $X=\mathbb{R}\times M$ and an adjoint-invariant inner product $\gamma$ on $\mathfrak{g}$. We use the trivial  connection on the principal bundle $P=X\times G\rightarrow X$. Employing the notations of Section \ref{sec:trivial_bundles}, we can write $\mathcal{L}(j^1s):=(\|\dot{\bar s}\|^2-\|\mathbf{d}\bar s\|^2)/2$ and the reduced Lagrangian density reads $\ell(\bar\sigma^1,\bar\sigma^2)=(\|\bar\sigma^1\|^2-\|\bar\sigma^2\|^2)/2$.

The corresponding instantaneous Lagrangian $L^\mathcal{L}$ and its reduced expression $l^\mathcal{L}$ are
\[ 
L^\mathcal{L}(\chi,\dot\chi,\gamma)=\frac{1}{2}\int_M\left(\|\dot\chi\|^2-\|\mathbf{d}\chi-\chi\gamma\|^2\right)\mu_M,\quad
l^ \mathcal{L}(\nu,\gamma)=\frac{1}{2}\int_M\left(\|\nu\|^2-\|\gamma\|^2\right)\mu_M.
\]
Thus we have recovered the spin glasses Lagrangian $l^\mathcal{L}$ considered in \cite{Dz1980}. The motion equations can be obtained either from $L^\mathcal{L}$ by dynamic reduction, or from $\mathcal{L}$ by covariant reduction. The covariant and dynamic reduced equations are
\[
\dot{\bar{\sigma}}^1+\operatorname{div}\bar\sigma^2=-\operatorname{Tr}\left(\operatorname{ad}^*_{\bar\sigma^2}\bar\sigma^2\right)\quad\text{and}\quad \dot\nu=\operatorname{div}^\gamma\gamma,
\]
respectively. One can directly see that these equations are equivalent by recalling the relations
\[
\bar\sigma^1_t=\nu_t\quad\bar\sigma^2_t=-\gamma_t,
\]
between the covariant and dynamic reduced variables. Recall that the compatibility condition for the first equation (obtained by covariant reduction) is equivalent to the advection equation $\dot{ \gamma} + \mathbf{d}^ \gamma \nu = 0 $ and $\mathbf{d}^ \gamma\gamma = 0 $ that need to be added to the second equation (obtained by dynamic reduction) to complete the system. All the considerations above can of course be generalized to arbitrary Lagrangians. In this case, the covariant reduced equation reads
\[
\frac{\partial}{\partial t}\frac{\delta \bar\ell}{\delta\bar{\sigma}^1}+\operatorname{div}\frac{\delta \bar\ell}{\delta\bar{\sigma}^2}=-\operatorname{ad}^*_{\bar\sigma^1}\frac{\delta \bar\ell}{\delta\bar{\sigma}^1}-\operatorname{Tr}\left(\operatorname{ad}^*_{\bar\sigma^2}\frac{\delta \bar\ell}{\delta\bar{\sigma}^2}\right),
\]
together with the compatibility condition. The dynamic reduced equation is
\[
\frac{\partial}{\partial t}\frac{\delta l^\mathcal{L}}{\delta\nu}=-\operatorname{ad}^*_\nu\frac{\delta l^\mathcal{L}}{\delta\nu}+\operatorname{div}^\gamma\frac{\delta l^\mathcal{L}}{\delta\gamma},
\]
together with the advection equation $\dot{ \gamma} + \mathbf{d}^ \gamma \nu = 0 $ and the vanishing of curvature $\mathbf{d}^ \gamma\gamma = 0 $.

If $M=\mathbb{R}^3$ and $G=SO(3)$, these systems of equations appear in the \textit{macroscopic description of spin glasses}, see equations (28) and (29) in \cite{Dz1980}. See also equations (3.9), (3.10) in \cite{IsKoPe1994}, or system (1) in \cite{Iv2000} and references therein and for an application to \textit{magnetic media} and the expression of more general Lagrangians. In this context, the variable $\nu$ is interpreted as the \textit{infinitesimal spin rotation}, $\delta l ^ \mathcal{L}/ \delta\nu$ is the \textit{spin density}, and the curvature $\mathbf{d}^\gamma\gamma$ is the \textit{disclination density}.


\addcontentsline{toc}{section}{References}
\begin{thebibliography}{300}






\bibitem[Castrill\'on-L\'opez, Garc\'{\i}a P\'erez, and Ratiu(2001)]{CaGaRa2001}
Castrill\'on-L\'opez, M., P.~L. Garc\'{\i}a P\'erez, and T.~S. Ratiu [2001], Euler-Poincar\'e reduction on principal bundles,  \textit{Lett. Math. Phys.}, \textbf{58}(2), 167--180. 

\bibitem[Castrill\'on-L\'opez and Marsden(2008)]{CaMa2008}
Castrill\'on-L\'opez, M. and J.~E. Marsden [2008],
Covariant and dynamical reduction for principal bundle field theories,
\textit{Ann. Glob. Anal. Geom.}, \textbf{34}(3), 263--285.

\bibitem[Castrill\'on-L\'opez and Ratiu(2003)]{CaRa2003}
Castrill\'on-L\'opez, M. and T.~S. Ratiu [2003],
Reduction in principal bundles: covariant Lagrange-Poincar\'e equations, \textit{Comm. Math. Phys.},  \textbf{236}(2), 223--250.

\bibitem[Castrill\'on-L\'opez, Ratiu, and Shkoller(2000)]{CaRaSh2000}
Castrill\'on-L\'opez, M., T.~S. Ratiu, and S. Shkoller [2000], Reduction in principal fiber bundles: covariant Euler-Poincar\'e equations, \textit{Proc. Amer. Math. Soc.}, \textbf{128}, 2155--2164.

\bibitem[Cendra, Marsden, and Ratiu(2001)]{CeMaRa2001}
Cendra, H., J.~E. Marsden, and T.~S. Ratiu [2001], Lagrangian reduction by stages, \textit{Mem. Amer. Math. Soc.}, \textbf{152}(722).




\bibitem[Dzyaloshinski\u\i(1980)]{Dz1980}
Dzyaloshinski\u\i, I.~E. [1980], Macroscopic description of spin glasses. \textit{Modern trends in the theory of condensed matter (Proc. Sixteenth Karpacz Winter School Theoret. Phys., Karpacz, 1979)}, \textit{Lecture Notes in Physics}, \textbf{115}, 204--224.






\bibitem[Gay-Balmaz and Ratiu(2008a)]{GBRa2008a}
Gay-Balmaz F. and T.~S. Ratiu [2008a], Reduced Lagrangian and Hamiltonian formulations of Euler-Yang-Mills fluids, \textit{J. Symp. Geom.}, \textbf{6}(2), 189--237.

\bibitem[Gay-Balmaz and Ratiu(2008b)]{GBRa2008b}
Gay-Balmaz F. and T.~S. Ratiu [2008b], The geometric structure of complex fluids, \textit{Adv. in Appl. Math.}, \textbf{42}(2) 176--275. 


\bibitem[Gotay, Isenberg, Marsden, Montgomery(2004)]{GoIsMaMo2004}
Gotay, M., J. Isenberg, J.~E. Marsden, and R. Montgomery [2004],
Momentum Maps and Classical Relativistic Fields. Part I: Covariant Field Theory, preprint, arXiv:physics/9801019v2, 2004

\bibitem[Gotay, Isenberg, Marsden(2004)]{GoIsMa2004}
Gotay, M., J. Isenberg, and J.~E. Marsden [2004],
Momentum Maps and Classical Relativistic Fields. Part II: Canonical Analysis of Field Theories, preprint, arXiv:math-ph/0411032v1, 2004





\bibitem[Isaev, Kovalevskii, and Peletminskii(1994)]{IsKoPe1994}
Isaev, A.~A.,  M.~Yu. Kovalevskii, and S.~V. Peletminskii [1994], On dynamics of various magnetically ordered structures, \textit{The Physics of Metals and Metallography} \textbf{77} (4),  342--347.

\bibitem[Ivanchenko(2000)]{Iv2000}
Ivanchenko, E.~A. [2000], Backward electromagnetic waves in a magnetically disordered dielectric, \textit{Low. Temp. Phys.} \textbf{26}(6), 422--424.





























\end{thebibliography}


{\footnotesize

\bibliographystyle{new}

}

\end{document}